\documentclass[%
 aip,
 jmp,%
 amsmath,amssymb,
 reprint,%
author-year,
author-numerical,
]{revtex4-1}

\usepackage{graphicx}
\usepackage{dcolumn}
\usepackage{bm}
\usepackage{caption}
\usepackage{subcaption}
\usepackage{pbox}
\usepackage{array}
\usepackage{makecell}
\usepackage{natbib}
\usepackage{bibentry}

\begin{document}


\title[Predicting the Spin Seebeck Voltage in Spin-polarized Materials: A Quantum Mechanical Transport Model Approach]{Predicting the Spin Seebeck Voltage in Spin-polarized Materials: A Quantum Mechanical Transport Model Approach} 

\author{Anveeksh Koneru}
 \altaffiliation[Ph.D. ]{Mechanical Engineering, West Virginia University.}
\author{Terence D. Musho}%
 \altaffiliation{Corresponding author: terence.musho@mail.wvu.edu.}
\affiliation{ 
Mechanical Engineering, West Virginia University, Morgantown, WV
}%


\date{\today}

\begin{abstract}
The spin Seebeck effect has recently been demonstrated as a viable method of direct energy conversion that has potential to outperform energy conversion from the conventional Seebeck effect. In this study, a computational transport model is developed and validated that predicts the spin Seebeck voltage in spin-polarized materials using material parameter obtain from first principle ground state density functional calculations. The transport model developed is based on a 1D effective mass description coupled with a microscopic inverse spin Hall relationship. The model can predict both the spin current and voltage generated in a non-magnetic material placed on top of a ferromagnetic material in a transverse spin Seebeck configuration. The model is validated and verified with available experimental data of La:YIG. Future applications of this model include the high-throughput exploration of new spin-based thermoelectric materials.
\end{abstract}

\pacs{Valid PACS appear here}
\keywords{Thermoelectric, Spin Seebeck Effect, NEGF}
\maketitle


\section{\label{sec:level1}Introduction}

Tougher sanctions on fossil fuel emissions and greater energy demand around the globe is forcing us to rely more on renewable energy resources. Clean energy technologies like thermoelectric power generation provides one solution to ease our dependence on non-renewable energy resources. However, the efficiency of thermoelectrics has been limited due to the inherent coupling of the electronic and thermal carriers. Many different approaches like grain boundary scattering~\citep{ZHAO2008259}, band structure engineering~\citep{C3EE43099E}$^,$~\citep{peiyanzhongTE}, substitutional effects~\citep{Chen1r} were incorporated to improve the efficiency of thermoelectrics. Though these methods improved the performance to certain extent, the phonon and electron interactions still hamper the commercial applicability of thermoelectrics. More recently, an avenue to decouple these interactions has been experimental demonstrated using temperature gradient induced spin currents~\citep{spintronics:1c}$^,$~\citep{spincaloritronics:1f}$^,$~\citep{Observation:1g}$^,$~\citep{spinsemiconducting:1l}$^,$~\citep{spinsemiconducting:1m}$^,$~\citep{spincaloritronics:1x}. With this new discovery comes the need for new transport models to understand and optimize their response. In providing a solution, this research is focused on the both the development of a 1D spin-transport model and validating the model using the available experimental data.

Conventional thermoelectric energy conversion utilizes the principle of Seebeck effect~\citep{Seebeck1} to convert thermal gradient into electric voltage. In this effect, the energy conversion takes place when majority charge carriers drift away from the region of high temperature. A new approach to design thermoelectric modules relies on a slightly different principle involving the electron's spin. The pioneering research in 2008 by Uchida et.al.~\citep{Observation:1g} has opened a new avenue to extract additional heat energy by utilizing the intrinsic angular momentum of electrons, colloquially known as spin. This field that explores charge, spin and energy transport due to temperature difference is called spin caloritronics~\citep{spincaloritronics:1f}$^,$~\citep{SLONCZEWSKI1996} and is a transpiring field with immense potential in heat conversion applications.

Using a principle called spin Seebeck effect a spin voltage can be generated in a metal contact attached on a ferromagnetic material (FM) due to a thermal gradient in the FM material lattice. There are two configurations reported in literature to extract the spin voltage, longitudinal~\citep{0953uchida} and transverse~\citep{UCHIDA2010524}$^,$~\citep{Observation:1g} configurations. The Figure~\ref{fig:three_probe_system}, shows a transverse configuration arrangement, which is the configuration of interest in this research, to extract spin voltage from a ferromagnetic material. Due to a thermal gradient, spin current is transferred from the FM substrate into the attached nonmagnetic metal (NM) contact. Due to the high spin orbital coupling associated with heavy nonmagnetic metals such as Pt~\citep{6516040}$^,$~\citep{spinorbital}, the spin carriers are separated and accumulated on the two ends of the metal. In response, a voltage is generated along the transverse direction proportional to the amount of accumulated spin carriers. This effect is opposite to the Hall effect where an applied electric field in the presence of a magnetic field separates the charge carriers while here on the contrary the separation of spin carriers generates the voltage hence named as the inverse spin Hall effect. Only spin polarized or ferromagnetic materials have the capability to generate spin currents and transfer into the Pt metal contacts through the inverse spin Hall effect. Hence, this research is primarily focused on studying spin-polarized materials.

In order to theoretically predict the inverse spin Hall voltage (V$_{ISHE}$) of spin polarized materials, we developed a 1-D spin transport model by combining non equilibrium Green's function formalism (NEGF)~\citep{DATTA2000253} and spin transport theory~\citep{Xiao:1y1}. In this approach, first the fundamental parameters of a material such as lattice constant, band gap, Fermi energy, effective mass and magnetization of the material lattice, were calculated using density functional theory (DFT). The NEGF model used this information to calculate the surface current for the spin channels independently and spin transport theory will then be used to calculate the spin current injection and inverse spin Hall voltage generated in the attached metal contact.

NEGF modeling has been implemented in various studies to describe the quantum transport in different materials in the presence of thermal bias~\citep{negfDFT11}$^,$~\citep{PhysRevBspinnegf}. There also available softwares like TranSIESTA~\citep{PhysRevBtrans} that incorporate the application of this formalism. A combination of DFT and NEGF~\citep{negfDFT11}; or NEGF and spin transport theory~\citep{PhysRevBspinnegf}, were used to describe the electron transport but a combination of DFT+NEGF+spin transport theory to calculate the V$_{ISHE}$ in magnetic materials is near to less in literature. Hence, this research article aims to develop a 1-D model using a combination of NEGF formalism and spin transport theory by incorporating the parameters obtained from DFT calculations.

Transverse spin Seebeck configuration, Figure~\ref{fig:three_probe_system}, which is the major scope of this research, has been experimentally verified using La:YIG~\citep{spinsemiconducting:1m} and NiFe~\citep{UCHIDA2010524}. Insulating magnetic materials like La:YIG displaying spin Seebeck effect is a strong indication of magnaon driven spin Seebeck that is caused due to spin redistribution in a material and also proves the capability of spin carriers in a material lattice to generate voltage. In addition to this, semiconducting or insulating oxide magnetic materials have immense scope in this emerging field due to their ability to accommodate wide variety of substitutions and tune various electronic and magnetic properties. Hence we chose La:YIG as our material of study to validate and verify the developed model. LaY$_2$Fe$_5$O$_{12}$ (La:YIG) material lattice is obtained by substituting one Yttrium with Lanthanum in Y$_3$Fe$_5$O$_{12}$ (YIG). There have been few theoretical studies reporting the fundamental property study using DFT to calculate lattice parameter, electronic band structure and effective mass data for YIG~\citep{0295xingtao}$^,$~\citep{yigyongnian}$^,$~\citep{baettigyig} but there is no available electronic band structure data for La:YIG. Hence, first principle calculations based of DFT were performed on YIG to compare the effective mass and band gap data with literature available for YIG, and a similar approach has been applied to calculate the electronic band structure data for La:YIG.


\section{\label{sec:level2}1-D Green's function approach}
Using a DFT approach, a supercell arrangement or a system of isolated molecules can be solved for their electronic structure relaxation type solid state physics problems. In case of quasi-particle transport phenomena calculations across two boundaries, the effects of chemical potential due an external bias or variation of charge density in the scattering region while considering temperature effects cannot be solved using DFT alone to replicate the boundary conditions in a quantum transport. However, the combination of DFT with NEGF formalism is a powerful tool to study quantum transport phenomena in nanoscale region. The NEGF formalism is a self-consistent method, where Schr\"{o}dinger's equation and Poisson's equation are solved self-consistently for a copmosite effective mass system.

\subsection{\label{sec:sublevel2}Non-Equilibrium Green's Function (NEGF) formalism}

\begin{figure}[!tbp]
  \begin{subfigure}[b]{0.4\textwidth}
    \includegraphics[trim={0 2cm 0 8cm},width=1.05\columnwidth]{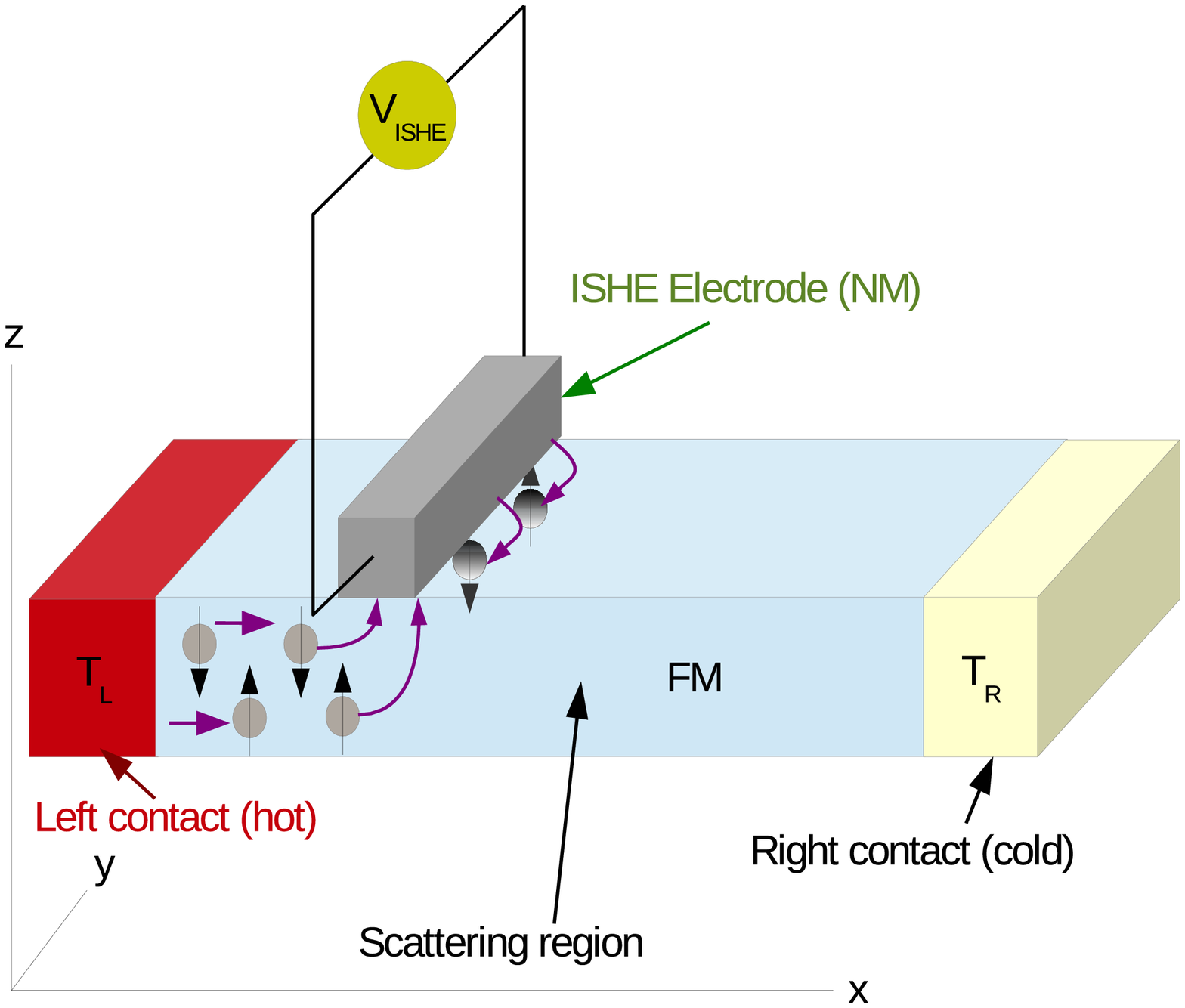}
    \caption{A three probe system consisting of left electrode with temperature T$_H$, right electrode with temperature T$_C$ and an ISHE electrode on the scattering region to measure the amount of current from the scattering region into the electrode.}
    \label{fig:three_probe_system}
  \end{subfigure}
  \begin{subfigure}[b]{0.4\textwidth}
    \includegraphics[trim={0 19cm 0 2cm},width=1.05\columnwidth]{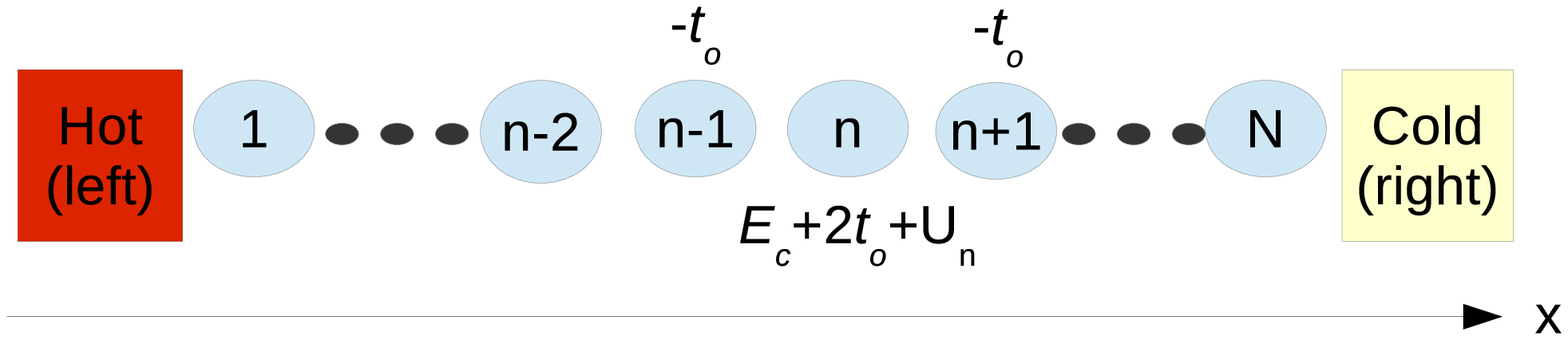}
    \caption{Description of the scattering region divided into N lattice points.}
    \label{fig:lattice}
  \end{subfigure}
  \caption{The number of grid points N depends on length of the lattice.}
    \label{fig:model}
\end{figure}

A composite system modeled in this research is shown in Figure~\ref{fig:three_probe_system} which is divided into four regions, i.e., the left contact, the scattering region, the right contact and the inverse spin Hall effect electrode (ISHE) which typically is Pt metal. The ISHE electrode is a conceptual floating probe used to calculate net spin flux flow between the FM and NM at various locations along the lattice. These probes are conceptually used to extract electrons from the device or inject into the device, at the region of study, to effectively calculate the scattering and transmission of electrons due to the applied temperature bias. The model incorporates the scattering effects that include connection of the channel to the contacts on the two ends and interactions withing the channel. To describe the system, two components that represent outflow [$\Sigma^{out}$]\{$\psi$\} and inflow \{s\} from the contacts should be added to the usual time-independent Schr\"{o}dinger equation represented by E\{$\psi$\} = [H]\{$\psi$\} which is given by the Equation~\ref{eq:system_sch}.

\begin{equation}
  E\{\psi\} = [H]\{\psi\} + [\Sigma^{out}]\{\psi\} + \{s\}
  \label{eq:system_sch}
\end{equation}
where $\psi$ is the many-particle wave function between the contacts.

From the above, \{$\psi$\} can be written as the following,
\begin{equation}
  \{\psi\}  =[G]\{s\}.
    \label{eq:psi_1}
\end{equation}

Where [G] is,
\begin{align}
  [G]  = [EI - H - \Sigma^{out}]^{-1}, \nonumber\\
\text{replacing} \quad \Sigma^{out}= \Sigma^{out}_l-\Sigma^{out}_r, \nonumber\\
  [G]  = [EI - H - \Sigma^{out}_l-\Sigma^{out}_r]^{-1}.  
    \label{eq:psi_2}
\end{align}
Here, the $\Sigma^{out}_l$ and $\Sigma^{out}_r$ terms are the outflow energies at the left and right contacts respectively. In addition to the outflow and inflow terms, the electrostatic potential energy (U) of the channel has to be included in Equation~\ref{eq:psi_2} which now becomes,
\begin{align}
  [G]  = [EI - H - U -\Sigma^{out}_l-\Sigma^{out}_r]^{-1}.  
    \label{eq:G}
\end{align}
The electron density in the scattering region which can be written as \{$\psi\}\{\psi\}^\dag$, generally represented by G$^n$, varies with the applied thermal bias. Using Equation~\ref{eq:psi_1}, G$^n$ can be written as  $G^n = \{\psi\} \{\psi\}^\dag =[G]\{s\}\{s\}^\dag[G]^\dag$. Where $\{s\}\{s\}^\dag$ is written as $\Sigma^{in}$ (the inflow from the source), G$^n$ can now be written as,
\begin{equation}
 G^n =[G] \Sigma^{in} [G]^\dag.
    \label{eq:psi_4}
\end{equation}

As there are contacts that interfere with the system, analytical solution cannot be obtained for the Equation~\ref{eq:psi_4}. One way to solve this equation is by expressing the material as finite discrete volumes with points at cell centers as shown in Figure~\ref{fig:lattice}, and representing those points through a matrix form. In the current research the proposed model will assume isotropic behavior at the cross-section of each lattice point along the scattering region, and assume a 1-dimensional model. At each lattice point (n) along the x-direction, the amount of spin current generated due to the difference in the two spin populations at that position can be calculated. This spin current is injected into the attached ISHE electrode at that position which is converted into spin voltage due to the principle of inverse spin Hall effect. In order to calculate the spin current at each lattice point, the population density can be calculated by applying NEGF model to solve Equation~\ref{eq:psi_4}.

The Equation~\ref{eq:psi_4} represents the fundamental equation behind NEGF formalism. The inflow energy from the contacts occur due to the non-equilibrium thermal difference at the hot and cold ends on the left and right contacts respectively and can be calculated using Equation~\ref{eq:sigma_in}. Calculating each of the terms on the right hand side of the Equation~\ref{eq:psi_4}, yields the electron density in the channel,
\begin{align}
  \Sigma^{in}_l=\Gamma_l*f_l, \nonumber\\
  \Sigma^{in}_r=\Gamma_r*f_r, \nonumber\\
  \Sigma^{in}=\Sigma^{in}_l+\Sigma^{in}_r,
    \label{eq:sigma_in}
\end{align}
where $f_l$ and $f_r$ are temperature-dependent Fermi-Dirac distribution of the continuous energy states $\epsilon$ at left and right contacts respectively. The Fermi function can be described by Equation~\ref{eq:fermi}. Here, T$_{l(r)}$ is the temperature of the contact at left(right) end and $\epsilon$ is the energy. It is through this Fermi-Dirac distribution that an explicit temperature bias is applied to the system.
\begin{equation}
  f_{l(r)}(\epsilon) = \frac{1}{1+{e^{(\epsilon - \mu_{l(r)})/(k_B T_{l(r)})]}}}
  \label{eq:fermi}
\end{equation}

When the temperature bias is applied to the scattering region by the end contacts, the conduction electrons tend to move away from the hot contact. The two spin channels (spin-up ($\uparrow$) and spin-down($\downarrow$)) associated to electrons move at different rates due to their different effective mass values. The effective mass of the $\uparrow$ and $\downarrow$ conduction electrons depends on the curvature of the respective conduction bands and can be calculated by taking the harmonic mean of the effective masses in the three reciprocal vectors directions in the Brillouin zone as given in Equation~\ref{eq:effmass}.
\begin{align}
  E_{\downarrow(\uparrow)}(k) = E_{\downarrow(\uparrow)cx}+\frac{\hbar^2*k_{x-\Gamma}^2}{2*m_{\downarrow(\uparrow)x}^*},\\
  E_{\downarrow(\uparrow)}(k) = E_{\downarrow(\uparrow)cy}+\frac{\hbar^2*k_{y-\Gamma}^2}{2*m_{\downarrow(\uparrow)y}^*},\\
  E_{\downarrow(\uparrow)}(k) = E_{\downarrow(\uparrow)cz}+\frac{\hbar^2*k_{z-\Gamma}^2}{2*m_{\downarrow(\uparrow)z}^*},\\
  m_{\downarrow(\uparrow)conduction}^{*}  = 3*[\frac{1}{m_{\downarrow(\uparrow)x}^*}+ \frac{1}{m_{\downarrow(\uparrow)y}^*}+\frac{1}{m_{\downarrow(\uparrow)z}^*}]^{-1},
\label{eq:effmass}
\end{align}
where k is the wavevector and E$_{\downarrow(\uparrow)cx}$, E$_{\downarrow(\uparrow)cy}$ and E$_{\downarrow(\uparrow)cz}$ are the conduction band edges in the x-$\Gamma$, y-$\Gamma$ and z-$\Gamma$ directions respectively of $\downarrow(\uparrow)$ electrons. Using the effective mass values of $\downarrow(\uparrow)$ electrons and the respective electronic band gaps, the amount of current generated in the scattering region due to the temperature bias between the contacts can be calculated.

Every electron has multiple energy levels available to accommodate their movement. Depending on the electron's eigen energy, charge transport dominates in certain energy bands which can be calculated from transmission function given by,
\begin{equation}
\Xi=Trace[\Gamma_lG \Gamma_rG^{\dag}],
  \label{eq:transmission}
\end{equation}
where $\Gamma_l$ and $\Gamma_r$ are NxN left and right contact anti-Hermitian matricies of $\Sigma^{out}_l$ and $\Sigma^{out}_r$ respectively that govern the inscattering and outscattering from contacts. All these parameters can be calculated from,
\begin{align}
\Sigma^{out}_l=
  \left( \begin{array}{cccccccccc}
    -t_o*e^{i*k*a} & 0& . &.&.&.& 0\\
    0 &. & .&.&.&.& 0 \\
      .& &.&&&&.\\
     .& & &.&&&.\\
     0&.&.&.&.& . & 0  
  \end{array} \right), \nonumber\\
  \Gamma_l=i*[\Sigma^{out}_l-{\Sigma^{out}_l}^\dag],\nonumber\\
  \Sigma^{out}_r=
  \left( \begin{array}{cccccccccc}
    0 & 0& . &.&.&.& 0\\
    0 &. & .&.&.&.& 0 \\
      .& &.&&&&.\\
     .& & &.&&&.\\
     0&.&.&.&.& . & -t_o*e^{i*k*a}  
  \end{array} \right), \nonumber\\
  \Gamma_r=i*[\Sigma^{out}_r-{\Sigma^{out}_r}^\dag],
  \label{eq:Sigma_source}
\end{align}
where k is cos$^{-1}$(1-$\frac{E}{2t_o}$). E is the energy level in the fine spectrum of energy bands, $-t_o*e^{i*k*a}$ t$_o$ is analytical wavefunction given by,
\begin{equation}
t_o = \frac{\hbar^2}{2*m_e* a^2 *q},
\label{eq:to}
\end{equation}
where a is the distance between cell centers. The distance between the cell centers is also correlated with the energy cut-off. m$_e$ is the conduction electron effective mass calculated from Equation~\ref{eq:effmass} , q is the charge of an electron, $\hbar$ is the reduced Planck's constant.

The transmission function ($\Xi$) of an energy band when multiplied with the Fermi functions, gives the conductance for each energy level. By integrating the quantum conductance of an electron in a fine spectrum of energy bands, the overall quantum conductance of the electron with that eigen energy can be obtained. Incorporating the same process for all the eigen energy states, the overall quantum conductance in the material lattice can be obtained. The quantum conductance multiplied with $\frac{q^2}{h}$ yields the current generated by a electron of one eigen energy. As given in Equation~\ref{eq:total_current}, taking the sum of total current for all eigen energy states, gives the net current in the scattering region.

\begin{equation}
I_{tot} = \sum_{\epsilon=E_{min}}^{\epsilon=E_{max}}\frac{q^2}{h}\Xi(\epsilon) (f_l-f_r)dE \\
  \label{eq:total_current}
\end{equation}

The domain is divided into N lattice points, as shown in Figure~\ref{fig:lattice}, can be described by an NxN matrix called the Hamiltonian matrix. The higher value of N gives better approximation by increasing the cut off energy while also taking a toll on the computational expense. A 1-D Hamiltonian matrix with N lattice points, [H]$_{N*N}$, can be written as given in Equation~\ref{eq:Hamiltonian}. The NxN Hamiltonian matrix has 'N' eigen values which are the eigen energy states of the scattering region. In Equation~\ref{eq:Hamiltonian}, E$_c$ is the conduction band edge that can be obtained from DFT calculations and t$_o$ is represented in Equation~\ref{eq:to}. It can be observed that the Hamiltonian matrix of the scattering region depends on the parameters obtained from DFT calculations and the fineness of the grid.\\
\begin{widetext}
\begin{equation}
[H]=
  \left( \begin{array}{cccccccccc}
    E_c+2t_o & -t_o & 0 & 0 &.&.&.&.&.& 0\\
    -t_o & E_c+2t_o & -t_o &0 &.&.&.&.&.& 0 \\
    0 & -t_o & E_c+2t_o & -t_o & 0 &.&.&.&.& 0 \\
      .& &.&&&&&&&.\\
     .& & &.&&&&&&.\\
     .& & & &.&&&&&.\\
     .& & & & &.&&&&.\\
     .& & & & & &.&&&.\\
     .&. &. &. & &0 & -t_o & E_c+2t_o & -t_o&0\\
     .&.&.&.&.&.& 0 & -t_o & E_c+2t_o & -t_o\\
     .&.&.&.&.&.&.& 0 & -t_o & E_c+2t_o 
  \end{array} \right)
  \label{eq:Hamiltonian}
\end{equation}
\end{widetext}

\begin{figure}[!ht]
\centering
\vspace{1cm}
\includegraphics[trim={0cm 6cm 3cm 4cm},scale=0.4]{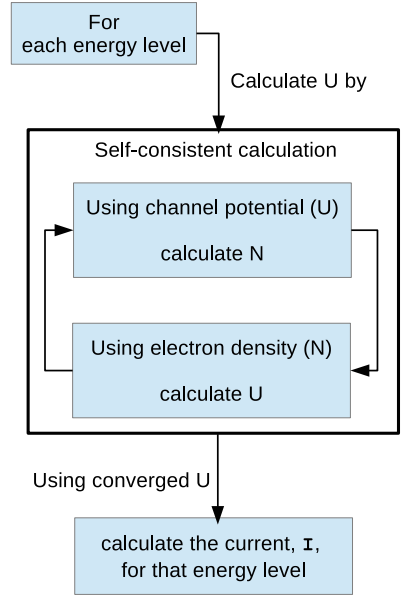}
\caption{NEGF flow chart of the self-consistent method. The self-consistent criteria is when consecutive energy difference reaches below 1e-6.}
\vspace{0cm}
\label{fig:NEGF_flow_chart}
\end{figure}

For each energy level, the potential of the scattering channel (U) and the electron density (N) can be calculated iteratively until self-consistency in the system is reached. This is described in the flow chart in Figure~\ref{fig:NEGF_flow_chart}. The self consistent procedure accounts for the electron correlation within each spin channel. After a converged self-consistent estimation of U, the electron density for that eigen energy state, which is the trace of $G^n$ matrix, is used to calculate the current generated at that state. As, isotropic conditions are assumed at each lattice point, trace of $G^n$ matrix in Equation~\ref{eq:psi_4} gives electron density per unit area.
Summing the currents over all eigen energy states gives the total current in the scattering region. Using the converged value of U at each energy level, the current in the channel at that respective energy level can be calculated. By summing the currents at all energy levels, the total current in the scattering region can be calculated.

In this research the spins will be treated independent of each other. This commonly is referred to as the Stoner model~\citep{PhysRevBstonermodel}. In doing so, the currents of the spin-up, I$_{\uparrow}$, and spin-down, I$_{\downarrow}$, can be obtained from Equation~\ref{eq:Itot_up}, for the respective $\Xi_{\uparrow}$ and $\Xi_{\downarrow}$ obtained from Green's formalism. In a ferromagnetic material there is a net spin-polarization present in the lattice. This can lead to spin-polarized charge current under temperature bias. The difference in the Fermi level of spin-up and spin-down conduction electrons in a material causes imbalance between the populations of spin-up and spin-down electrons that can lead to spin-polarized charge current.\\
\begin{align}
I_{tot(\uparrow)} = \sum_{\epsilon=E_{min}}^{\epsilon=E_{max}}\frac{q^2}{h}\Xi_{(\uparrow)}(\epsilon) (f_{l(\uparrow)}-f_{r(\uparrow)})dE \nonumber\\
I_{tot(\downarrow)} = \sum_{\epsilon=E_{min}}^{\epsilon=E_{max}}\frac{q^2}{h}\Xi_{(\downarrow)}(\epsilon) (f_{l(\downarrow)}-f_{r(\downarrow)})dE
  \label{eq:Itot_up}
\end{align}


\section{\label{sec:level3}Calculation of inverse spin Hall voltage}
\label{sec:Spin_Seebeck}

When a temperature gradient ($\nabla$T) is applied, the $\uparrow$($\downarrow)$ components move away from hot source at different rates and hence have different spin Seebeck coefficients, S$_{\uparrow(\downarrow)}$. The 1-D model in this research treats the two spin channels independently. To calculate the S$_{\uparrow(\downarrow)}$, a voltage in the reverse bias will be applied between the left and right electrodes ($V_{\uparrow(\downarrow)}$), see Figure~\ref{fig:lattice}. This will inject the spin current into the ferro-magnetic (FM) material (scattering region) that opposes the spin current generated in the FM due to temperature bias. This $V_{\uparrow(\downarrow)}$ will be iterated until the respective spin current ($I_{\uparrow(\downarrow)}$) flow is zero between the left contact and right contact. The respective slopes in the IV curve gives conductivity $\sigma_{\uparrow}$ and $\sigma_{\downarrow}$ of spin-up and spin-down electrons. The $V_{\uparrow}$ for which $I_{tot(\uparrow)}$, in Equation~\ref{eq:Itot_up}, is zero gives the Seebeck voltage of the $\uparrow$ channel and the spin Seebeck coefficient for $\uparrow$ channel, $S_{\uparrow}$, at that temperature would be S$_{\uparrow}$ = $\frac{V_{\uparrow}}{\Delta T}$. A similar iterative procedure can yield spin Seebeck coefficient for $\downarrow$ channel S$_{\downarrow}$. The electronic contribution to thermal conductivity $k_e$ can be calculated from Fourier's law, $k_e$ = $\frac{I_\uparrow + I_\downarrow}{A} \frac{\Delta x}{\Delta T}$.

The presence of temperature gradient along the ferromagnetic material changes the density flux of spin-up and spin-down electrons. Exploring the substitutional effects to improve the difference in this spin-up and spin-down density flux will make a material suitable for spin caloritronic applications. Sometimes the substitutions make the material spin-polarized while also making the material conductive which hampers the application of the material for thermoelectrics. Hence theoretical characterization of a material performance for spin caloritronic applications can be made using the developed model in this research.

\subsection{Spin Current Across the Magnetic and Non-Magnetic Interfaces}
\label{sec:Js across the interface }

At each location along the x axis, in Figure~\ref{fig:lattice}, different amount of spin current is injected into the NM contact from the FM material. In a conventional Seebeck effect, the temperature difference between the junctions of two dissimilar metal joints generates a net voltage. Likewise, in case of spin Seebeck effect (SSE), the difference in the temperature between the magnons (electron spin waves) in the ferromagnetic film (considered to be close to that of the FM temperature) and the electrons in the attached NM contact (considered to be close to that of the NM temperature), generates an inverse spin Hall voltage in the NM contact.

As SSE was observed even in magnetic insulators, the Spin Seebeck coefficient cannot be fully expressed as a sole function of conduction electrons. Hence, a microscopic theory proposed by Xiao et al ~\citep{Xiao:1y1} was employed in this research that utilizes the calculated spin density from the quantum mechanical model to predict the spin current and voltage in the contact. As in conventional Seebeck effect where the temperature difference between the left and right contacts drives the charge current, here the temperature difference between the FM (T$_F$) and NM (T$_N$) drives the spin current across the interface of FM and NM. The model approximates that the difference between the T$_F$ and T$_N$, ($\Delta T= T_F-T_N$), changes from a positive value at the hot end to a negative value at the cold end, flipping the sign at the center which in close approximation to the experimental findings~\citep{spinsemiconducting:1m}.

The spin pumping current from FM to NM occurs due to thermal non-equilibrium between the two, given by Equation~\ref{eqn:Jsp}. The thermal spin pumping current J$_{sp}$ from FM to NM is proportional to T$_F$ and spin current fluctuations J$_{f}$ from NM to FM is proportional to T$_N$, given by Johnson-Nyquist~\citep{PhysRevLett.95.016601}$^,$~\citep{PhysRevB.79.174415}. Only the real component of the J$_{sp}$ contributes to the net J$_s$ towards the NM because the imaginary component averages to zero. The real part of the spin current from FM into the NM as extracted from literature~\citep{Tserkovnyak:1n1} is given by,
\begin{equation}
  J_{sp}  = \frac{\hbar}{4\pi}[g_r\boldsymbol{m}\times \boldsymbol{\dot{m}}],
\label{eqn:Jsp}
\end{equation}
where $\hbar$ is the reduced Plank's constant, g$_r$ is the real part of spin mixing conductance, $\boldsymbol{m}$ is the unit vector that is parallel to the magnetization in the material, $\boldsymbol{\dot{m}}$ is the rate of change of magnetization due to thermally activated dynamics in magnetization in FM. To compensate the energy transfer from FM into NM in the form of spin, a fluctuating spin current, J$_{f}$, flows from NM to FM, as extracted from literature~\citep{PhysRevLett.95.016601}$^,$~\citep{PhysRevB.79.174415} is given by,
\begin{equation}
  J_{f}  = \frac{-M_sV}{\gamma}[\gamma \boldsymbol{m}\times \boldsymbol{h^\prime}],
\label{eqn:Jfl}
\end{equation}
where M$_s$ is the saturation magnetization in the FM, V is the FM volume, $\gamma$ is the electron gyro-magnetic ratio and $\boldsymbol{h^\prime}$ is the resultant magnetic field.

Hence, the non-equilibrium thermal difference between FM and NM causes the net spin current, J$_s$, from FM to NM in the z-direction given by J$_s$= J$_{sp}$-J$_{fl}$
\begin{equation}
  \boldsymbol{J}_{s}  = \frac{M_sV}{\gamma}[\alpha^\prime \langle{m_x\dot{m_y}-m_y\dot{m_x}\rangle}
  -\gamma \langle {m_xh_y^\prime-m_y h_x^\prime} \rangle],
\label{eqn:Js}
\end{equation}
where $\alpha^\prime$ is the damping enhancement due to spin pumping given by $\gamma\hbar g_r/4\pi M_sV $, the x and y subscripts of $\boldsymbol{m}$ and $\boldsymbol{\dot{m}}$ are the respective components along x and y axes respectively (see Figure~\ref{fig:three_probe_system}). Using mean square deviation $\langle {\boldsymbol{m} \boldsymbol{\dot{m}}} \rangle$ and $\langle {\boldsymbol{m} \boldsymbol{\dot{h}^\prime}}\rangle$ can be approximated by Equation~\ref{eqn:mmdot} and Equation~\ref{eqn:mhprime} respectively as extracted from Xiao et al~\citep{Xiao:1y1}.

\begin{equation}
  \langle{\dot{m}_i(t) m_j(0)}\rangle  = \frac{-\sigma_{sd}^2}{4\pi \alpha} \int [\chi_{ij}(\omega)-\chi_{ij}^*(\omega)]e^{i\omega t} d\omega
\label{eqn:mmdot}
\end{equation}
\begin{equation}
  \langle{{m}_i(t) h_j^\prime(0)}\rangle  = \frac{-\sigma_{sd}^{\prime^2}}{2\pi\gamma} \int \chi_{ij}(\omega)e^{i\omega t} d\omega
\label{eqn:mhprime}
\end{equation}
$\chi(\omega)$ is the transverse dynamic susceptibility matrix given by Equation~\ref{eqn:suscept},
\begin{equation}
  \chi(\omega) = \frac{1}{(\omega_o-i\alpha\omega)^2-\omega^2}
  \begin{bmatrix}
    \omega_o-i\alpha\omega       & -i\omega \\
    i\omega     & \omega_o-i\alpha\omega \\
\end{bmatrix},
\label{eqn:suscept}
\end{equation}
where $\omega_o$=$ \gamma H_{eff}$ is the FM resonance frequency. $H_{eff}$ is the external magnetic field applied along the direction of thermal bias that causes saturation magnetization in the material.

Hence, by plugging Equations~\ref{eqn:mhprime} and ~\ref{eqn:mmdot} into Equation~\ref{eqn:Js} and rearranging for the time-averaged spin current transferred from FM to NM in the z direction as given in literature~\citep{Xiao:1y1} will be,
\begin{equation}
  \langle {J_s} \rangle = \frac{\gamma\hbar g_r k_B}{2\pi M_s V} (T_{FM}-T_{NM}).
\label{eqn:Iznew}
\end{equation}

The factor $\frac{\gamma\hbar g_r k_B}{2\pi M_s V}$ is called interfacial spin Seebeck coefficient. The Spin Seebeck voltage can be calculated based on the spin Hall current in the NM that is generated due to inverse spin Hall effect~\citep{inversespinhallcite}. The DC spin Hall current in the attached NM (Pt in this case) along y-direction is,
\begin{equation}
  J_c(x) \boldsymbol{\hat{y}} = \theta_H \frac{2 q \langle {J_s} \rangle}{\hbar A} \boldsymbol{\hat{z}} \times \boldsymbol{\hat{x}},
\label{eqn:Jcx}
\end{equation}
where $\theta_H$ is the spin Hall angle of NM contact (Pt in this case).
The Inverse spin Hall voltage generated at location x is given by Equation~\ref{eqn:Vishe_1},
\begin{equation}
 V_{ISHE}(x) = \rho l J_c(x) = \frac{l J_c(x)}{\sigma},
\label{eqn:Vishe_1}
\end{equation}
where $\rho$ is the electrical resistivity of Pt contact, $l$ is the length of the contact and J$_c$(x) is the spin current from FM to NM as calculated in Equation~\ref{eqn:Jcx}. 


\section{\label{sec:results}Results}

Spin Seebeck theory and NEGF transport theory combined with DFT can be used to theoretically calculate the spin voltage in the attached NM metal on the scattering region. As a model has been developed in this research, validating this model with the available experimental data is necessary.

\subsection{\label{sec:comp_details}Computational Details}
\begin{figure}[!ht]
  \begin{subfigure}[t]{0.49\textwidth}
    \includegraphics[trim={0cm 0cm 0cm 0cm},clip,width=0.9\columnwidth]{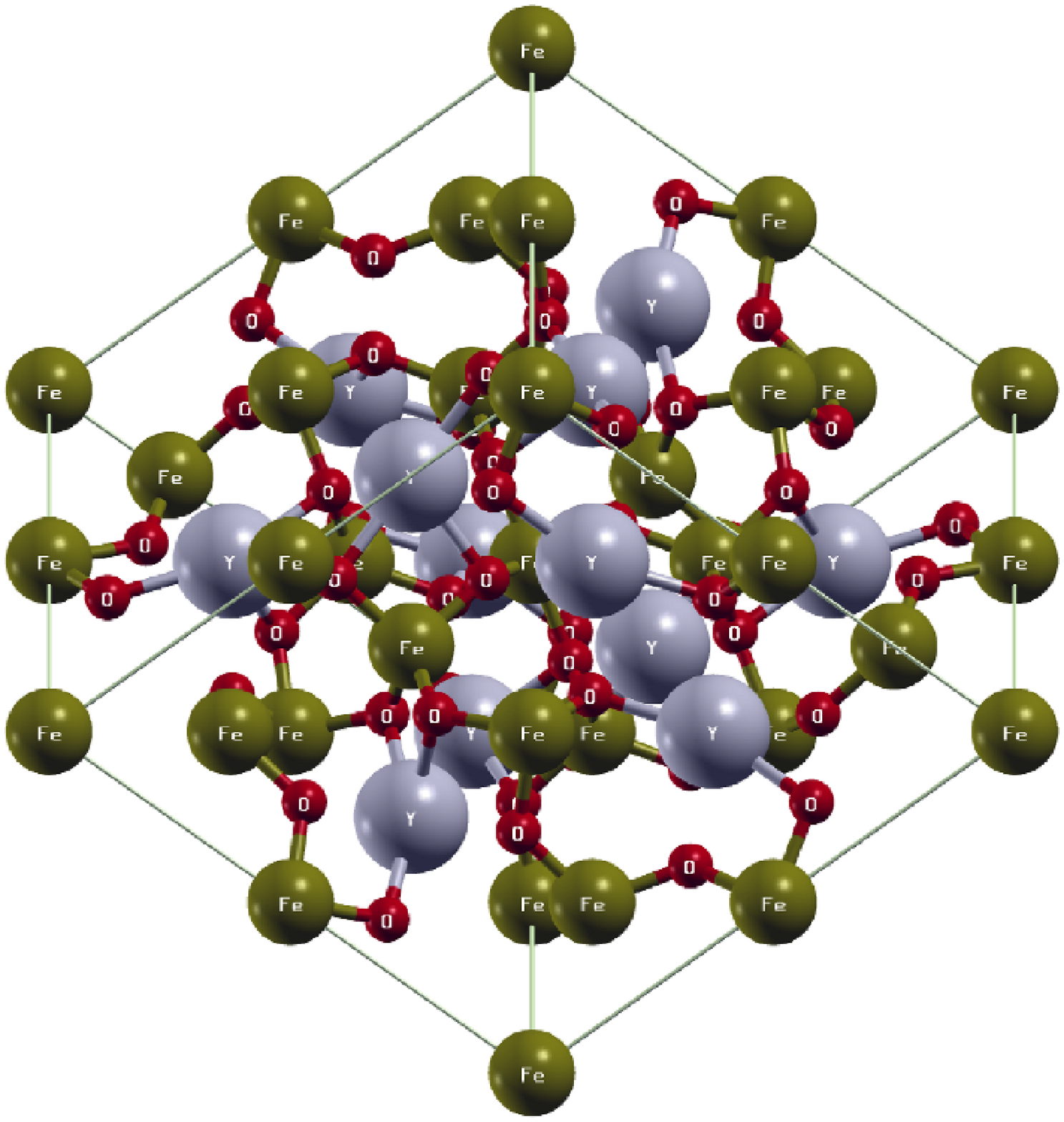}
    \caption{}
    \label{fig:yig_cell}
  \end{subfigure}
  \begin{subfigure}[t]{0.49\textwidth}
    \includegraphics[trim={0cm 0cm 0cm 0cm},clip,width=0.9\columnwidth]{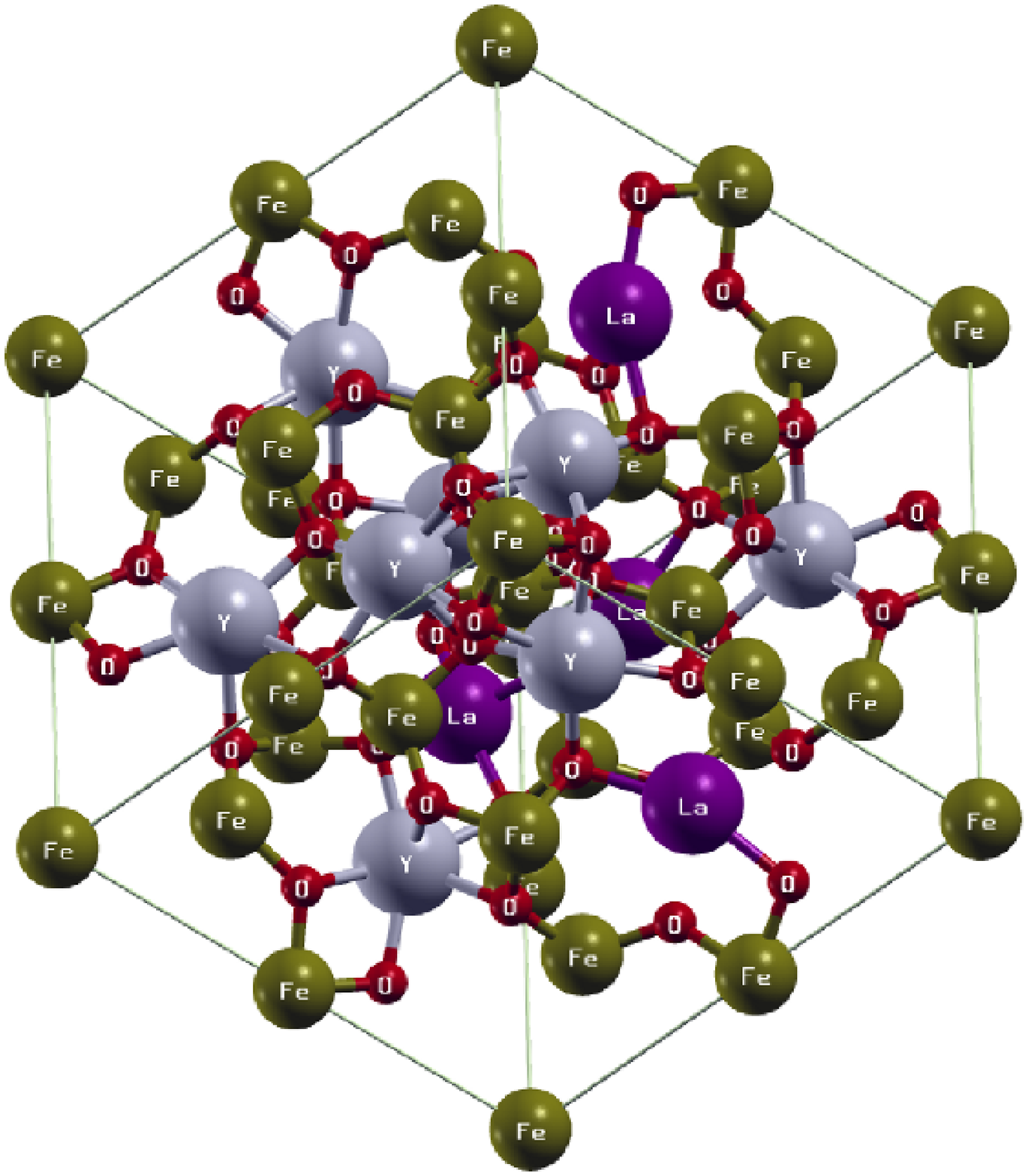}
     \caption{}
    \label{fig:UC_LaYIG}
  \end{subfigure}
  \caption{(a) A 80 atom super cell of YIG. The green balls represent Fe$^{+3}$ ions, the gray balls represent Y$^{+3}$ ions and the red balls represent O-2 ions.(b) The super cell of La:YIG. The four purple balls in La:YIG super cell that replace the Y$^{+3}$ ions represent La$^{+3}$ ions.}
    \label{fig:unitcell}
\end{figure}

Ab-initio calculations of the electronic band structure and structural properties were performed based on density functional theory using the plane wave scheme as implemented in Quantum Espresso package~\citep{0953-8984-21-39-395502}. In all the calculations the plane wave energy cut-off of 1220 eV was used to yield high convergence in energies. Super cells were relaxed to less than 5KPa and relative energy convergence of 10$^{-9}$ eV. The computed lattice constant of YIG of 11.62 $\AA$ matches well with the available experimental data of 12.39 $\AA$~\citep{baettigyig}. The exchange and correlation energy was described by the GGA as presented by PBE functional~\citep{Perdew:19} (QE-PBE). All the calculations were spin-polarized and the atomic cores were described by ultrasoft pseudo potentials~\citep{PhysRevB.41.7892}, as they efficiently handle localised electrons and provide accurate results comparable to all-electron calculations. The atomic valance of 4s$^1$3d$^7$, 5s$^2$4d$^1$ and 2s$^2$2p$^4$ was used for Fe, Y and O in the respective pseudopotentials. Calculations were performed on 80 atom super cell shown in Figure~\ref{fig:yig_cell} using a 4x4x4 Monkhorst-Pack k-point grid.

\subsection{\label{sec:YIG_data}YIG Model Results}
A primitive Y$_3$Fe$_5$O$_{12}$ unit cell has 20 atoms. A 2x2x1 super cell, Figure~\ref{fig:yig_cell}, was constructed that comprises 80 atoms. The Fe atoms with tetragonal and octahedral coordination carry majority of the magnetic moment of YIG while Y atoms having dodecahedral coordination and O atoms have little contribution to the magnetic moment. Fe atoms have highest number of unpaired d electrons that contribute to magnetic moment in the material. The electronic structure of YIG was calculated by implementing the scheme discussed in the previous section.

\begin{figure*}[!tbp]
  \begin{subfigure}[b]{0.32\textwidth}
    \includegraphics[width=\textwidth]{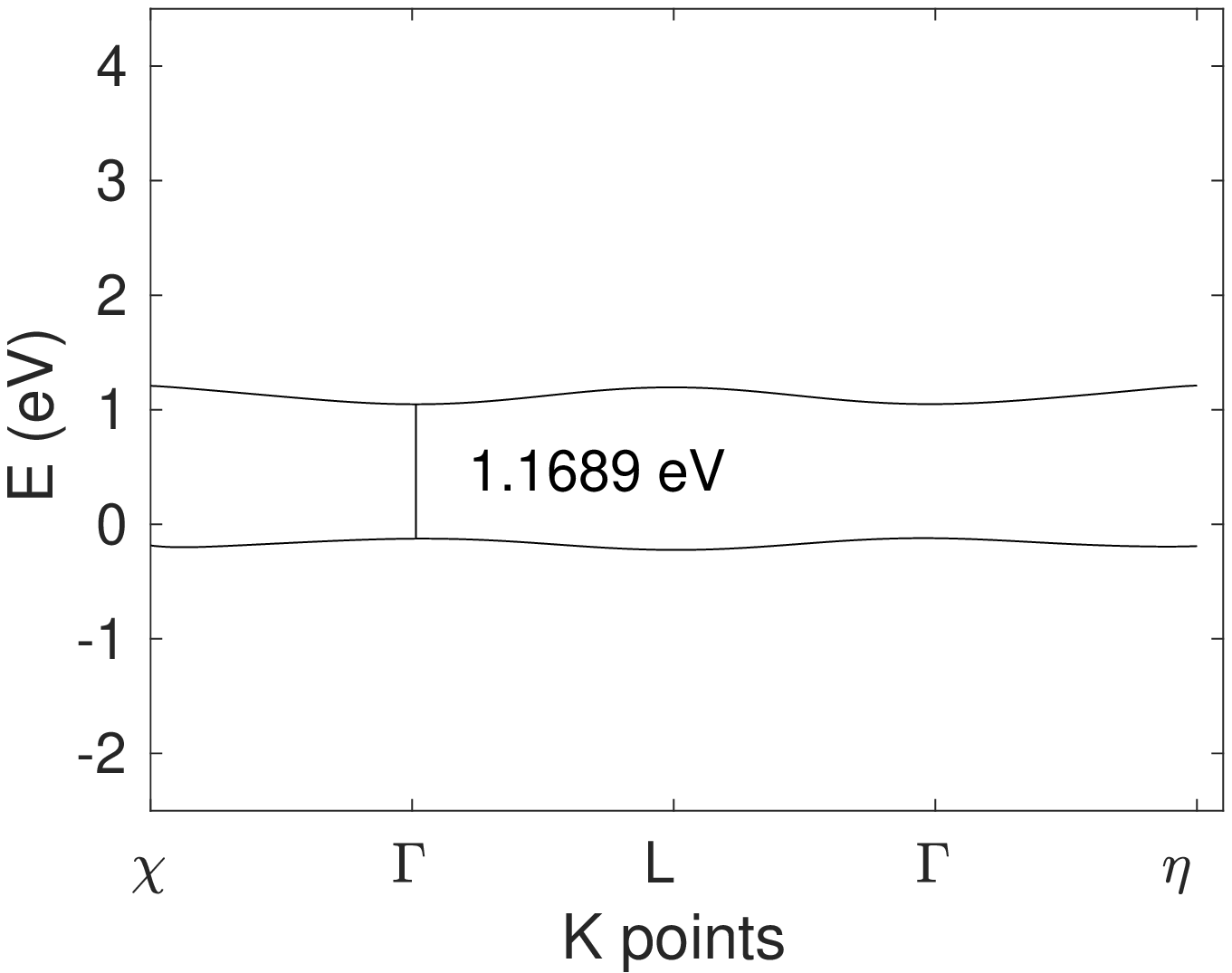}
    \caption{}
    \label{fig:f1}
  \end{subfigure}
  \begin{subfigure}[b]{0.32\textwidth}
    \includegraphics[width=\textwidth]{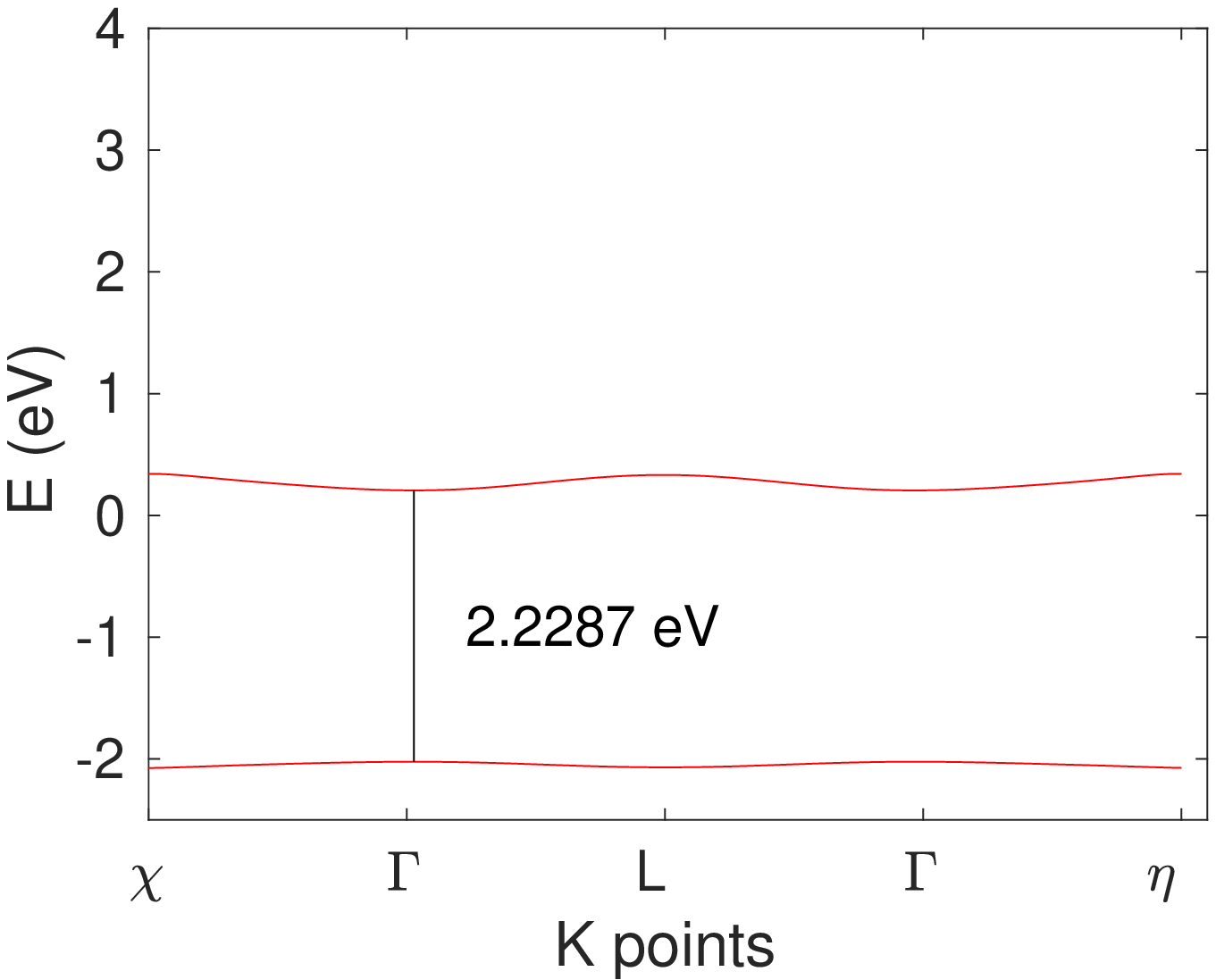}
     \caption{}
    \label{fig:f2}
  \end{subfigure}
  \begin{subfigure}[b]{0.32\textwidth}
    \includegraphics[width=\textwidth]{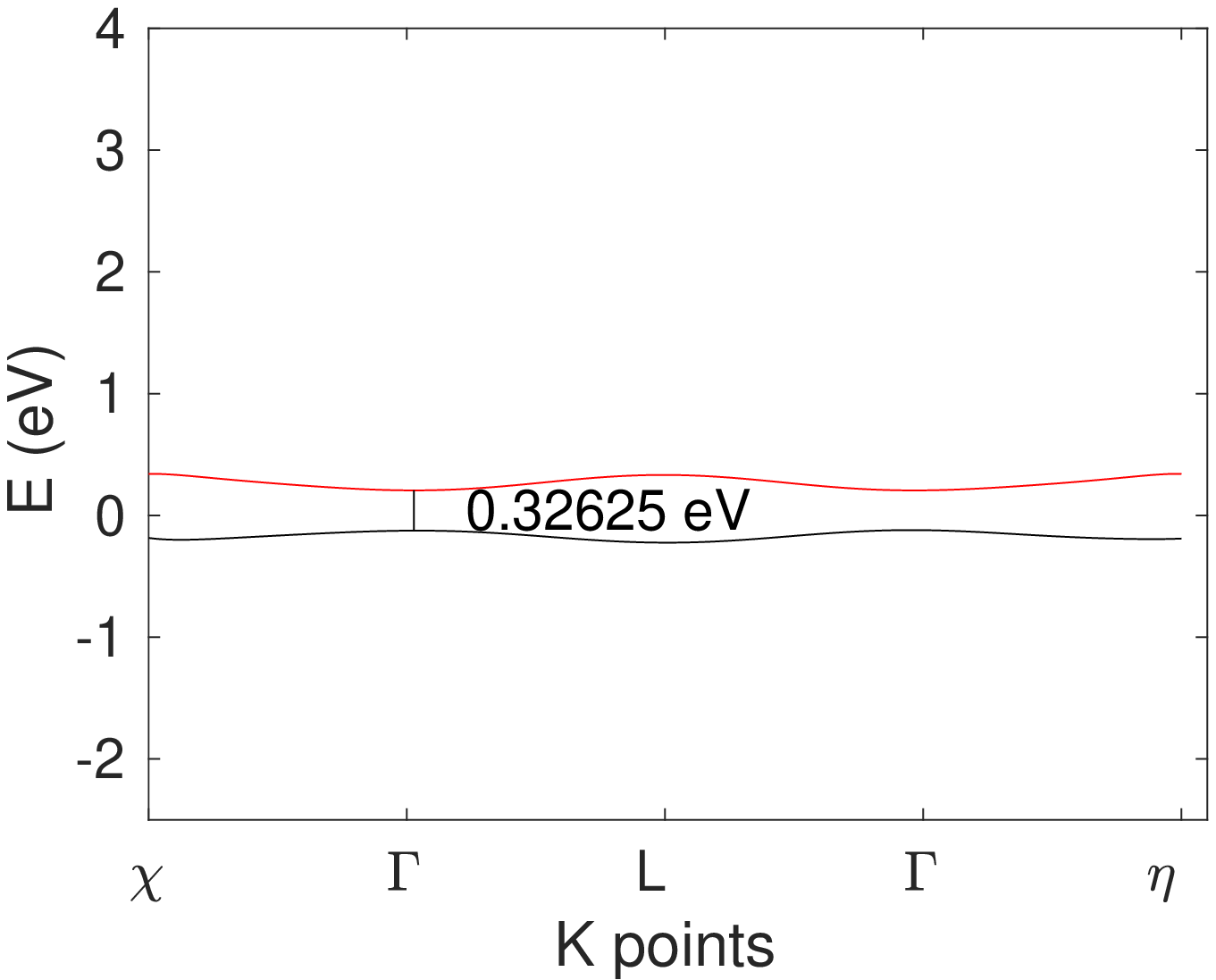}
    \caption{}
    \label{fig:f3}
  \end{subfigure}
  \hfill
  \begin{subfigure}[b]{0.32\textwidth}
    \includegraphics[width=\textwidth]{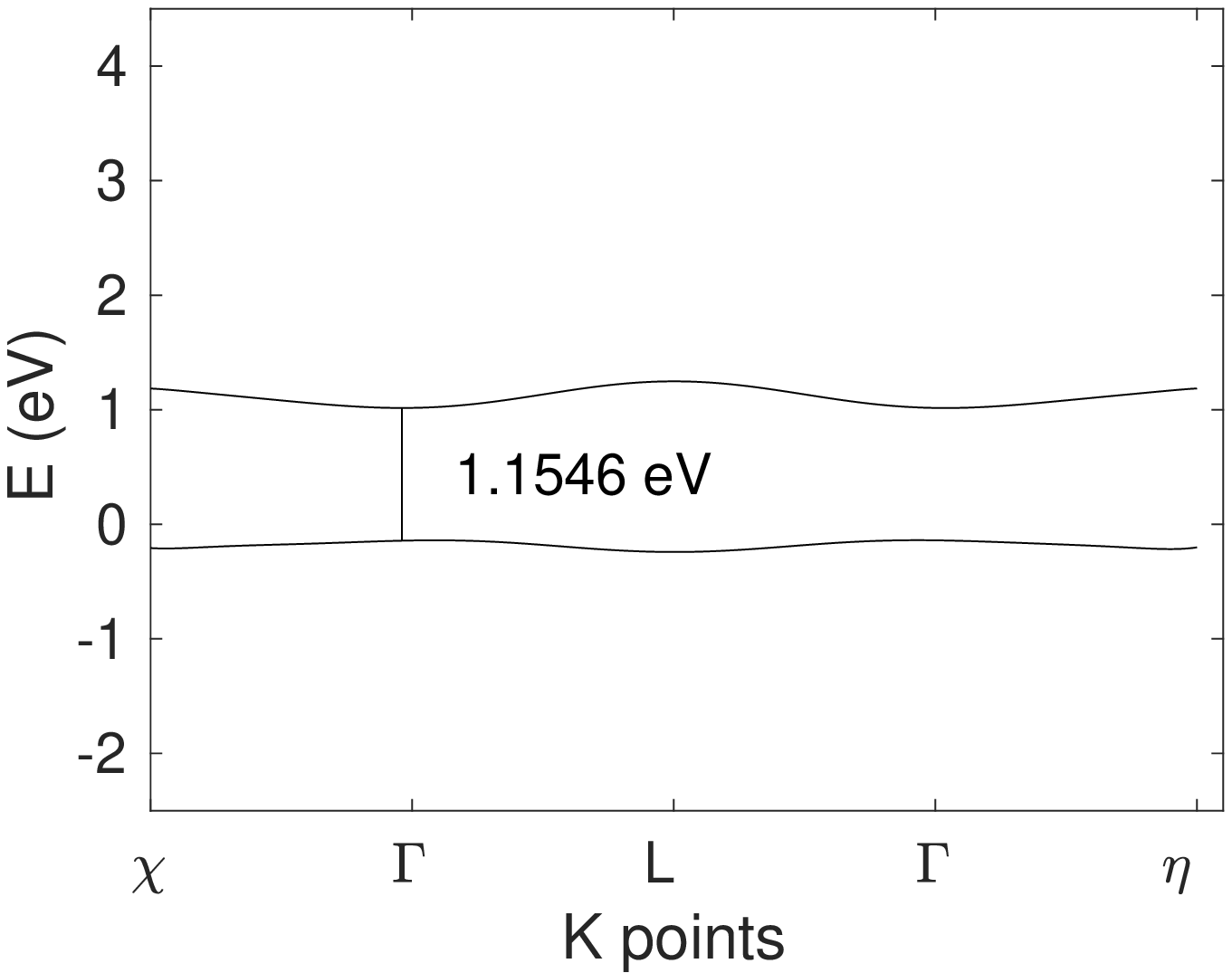}
    \caption{}
    \label{fig:laf1}
  \end{subfigure}
  \begin{subfigure}[b]{0.32\textwidth}
    \includegraphics[width=\textwidth]{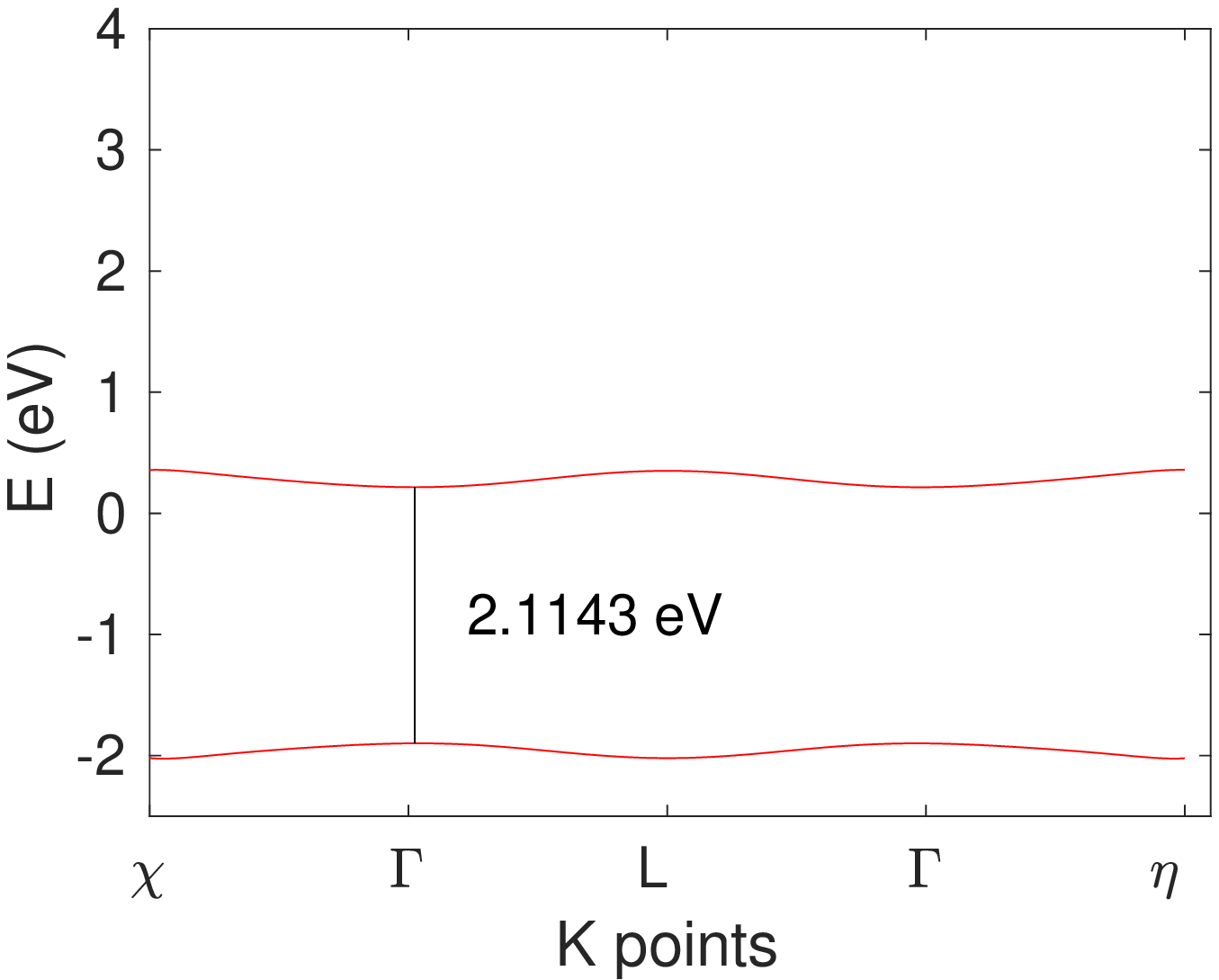}
    \caption{}
    \label{fig:laf2}
  \end{subfigure}
  \begin{subfigure}[b]{0.32\textwidth}
    \includegraphics[width=\textwidth]{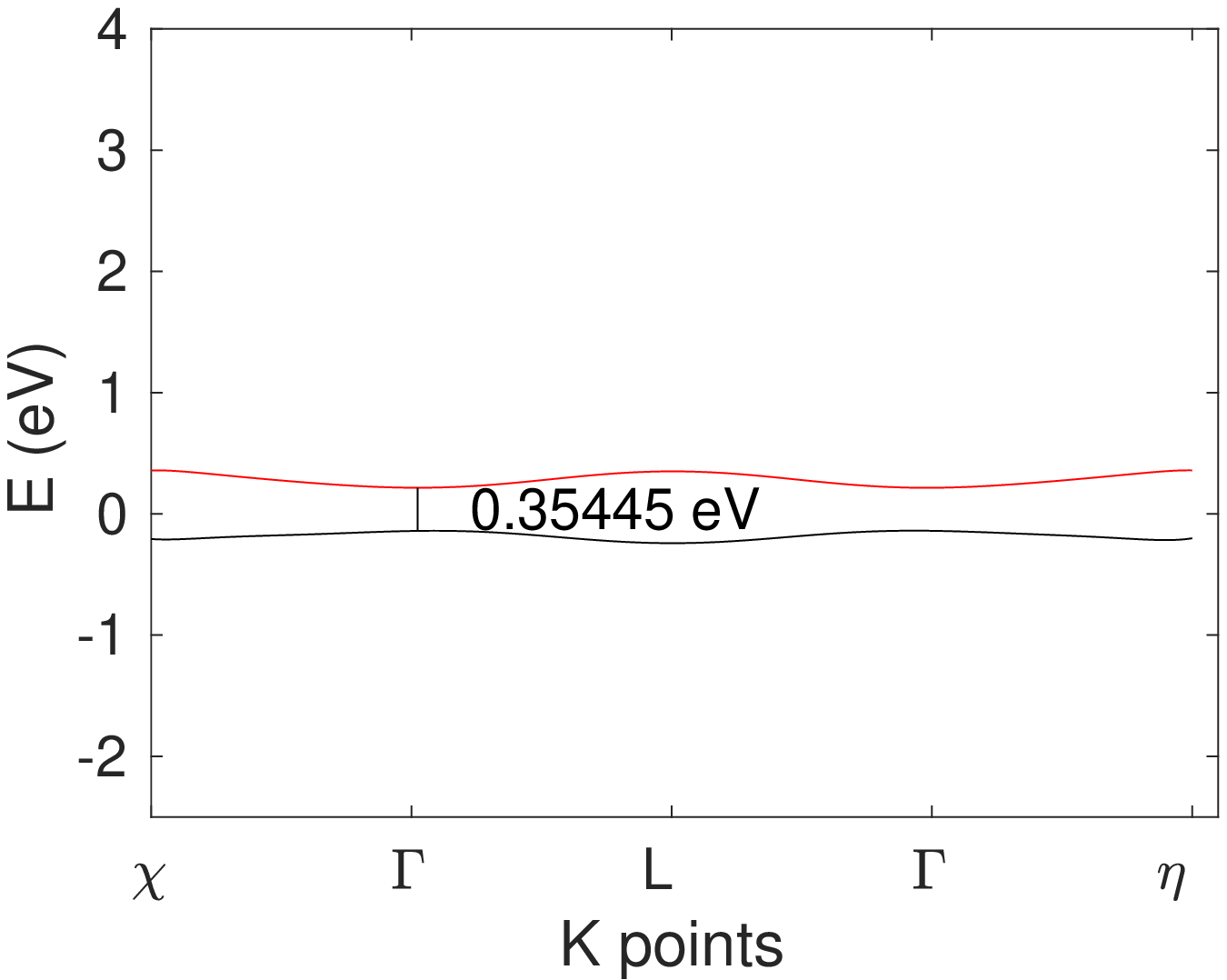}
    \caption{}
    \label{fig:laf3}
  \end{subfigure}
  \caption{(a),(b) and (c) respectively correspond to the spin up channel, spin down channel and combined valence-conduction bands of YIG material; while (d), (e) and (f) respectively correspond to the spin up channel, spin down channel and combined valence-conduction bands of La:YIG material at Fermi region. Compared to YIG, the band gap of individual channels decreased for La:YIG while the net band gap of La:YIG increased slightly.}
    \label{fig:YIG_bands}
\end{figure*}

The valence and conduction bands of spin up and spin down bands of YIG is shown in Figure~\ref{fig:f1} and Figure~\ref{fig:f2}. The spin down contribution to the bands is given in red, the spin up contribution to bands is given in black and the combined valence and conduction bands for the YIG material is shown in Figure~\ref{fig:f3}. The k-point path was chosen along the three reciprocal lattice vector directions. $\chi$(0,0,1/2), $\Gamma$(0,0,0), L(1/2,0,0) and $\eta$(0,-1/2,1/2) represent the symmetry k-points. $\chi$ to $\Gamma$, $\Gamma$ to L and $\Gamma$ to $\eta$ represent the three reciprocal vector directions. The $\uparrow$ and $\downarrow$ channels have a band gap of 1.1689eV and 2.2287eV respectively suggesting that spin up channel is more conductive than the spin down channel. In the combined bands of the YIG, the spin up channel's valance band and spin down channel's conduction band contribute to the majority of the charge transport. In the presence of thermal gradient, the conduction band that corresponds to the $\downarrow$ channel contributes to the electron movement creating a non-equilibrium in spin distribution along the lattice.

\begin{figure*}[!tbp]
  \begin{subfigure}[b]{0.32\textwidth}
    \includegraphics[width=\textwidth]{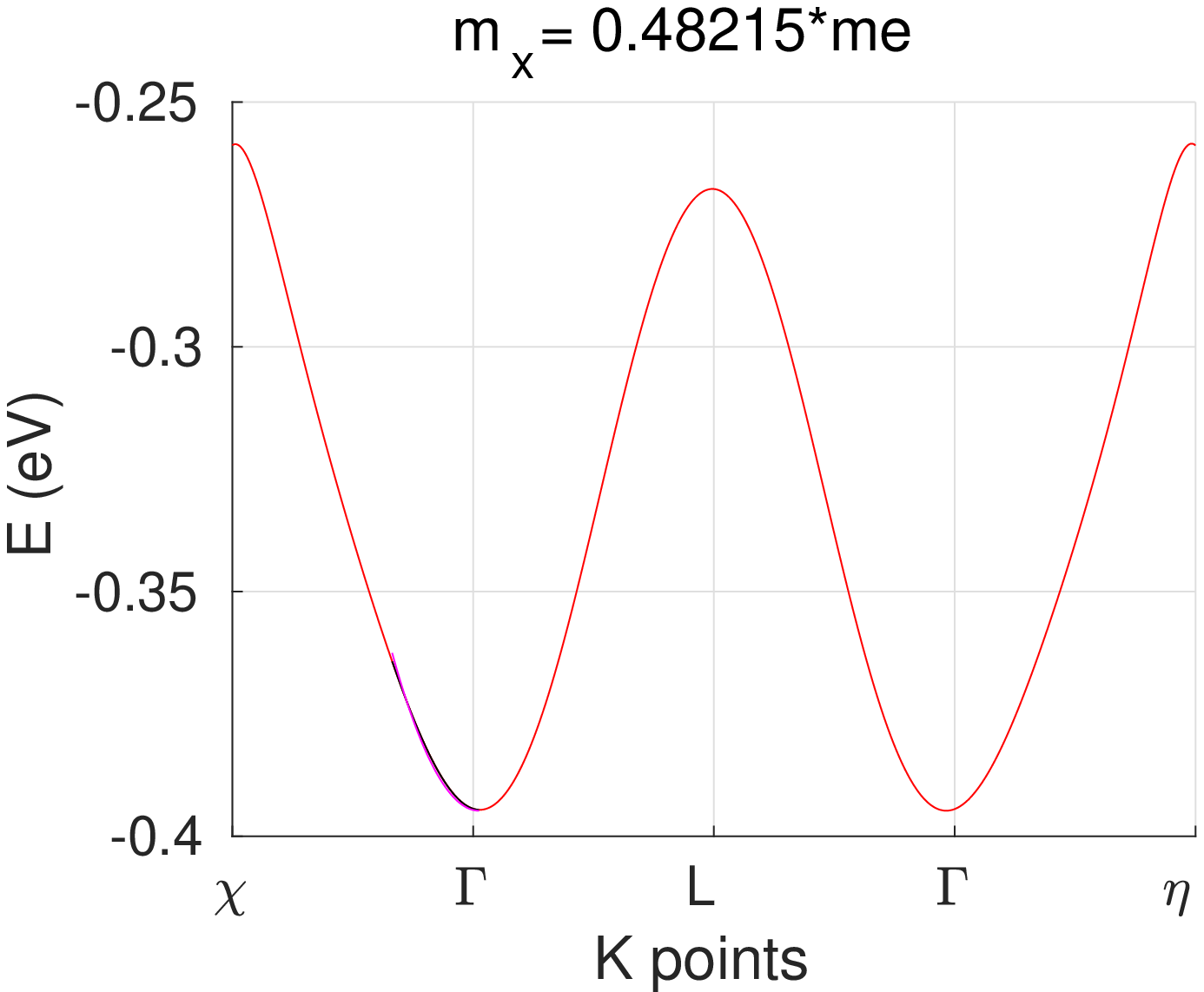}
    \caption{}
    \label{fig:eff_f1}
  \end{subfigure}
  \begin{subfigure}[b]{0.32\textwidth}
    \includegraphics[width=\textwidth]{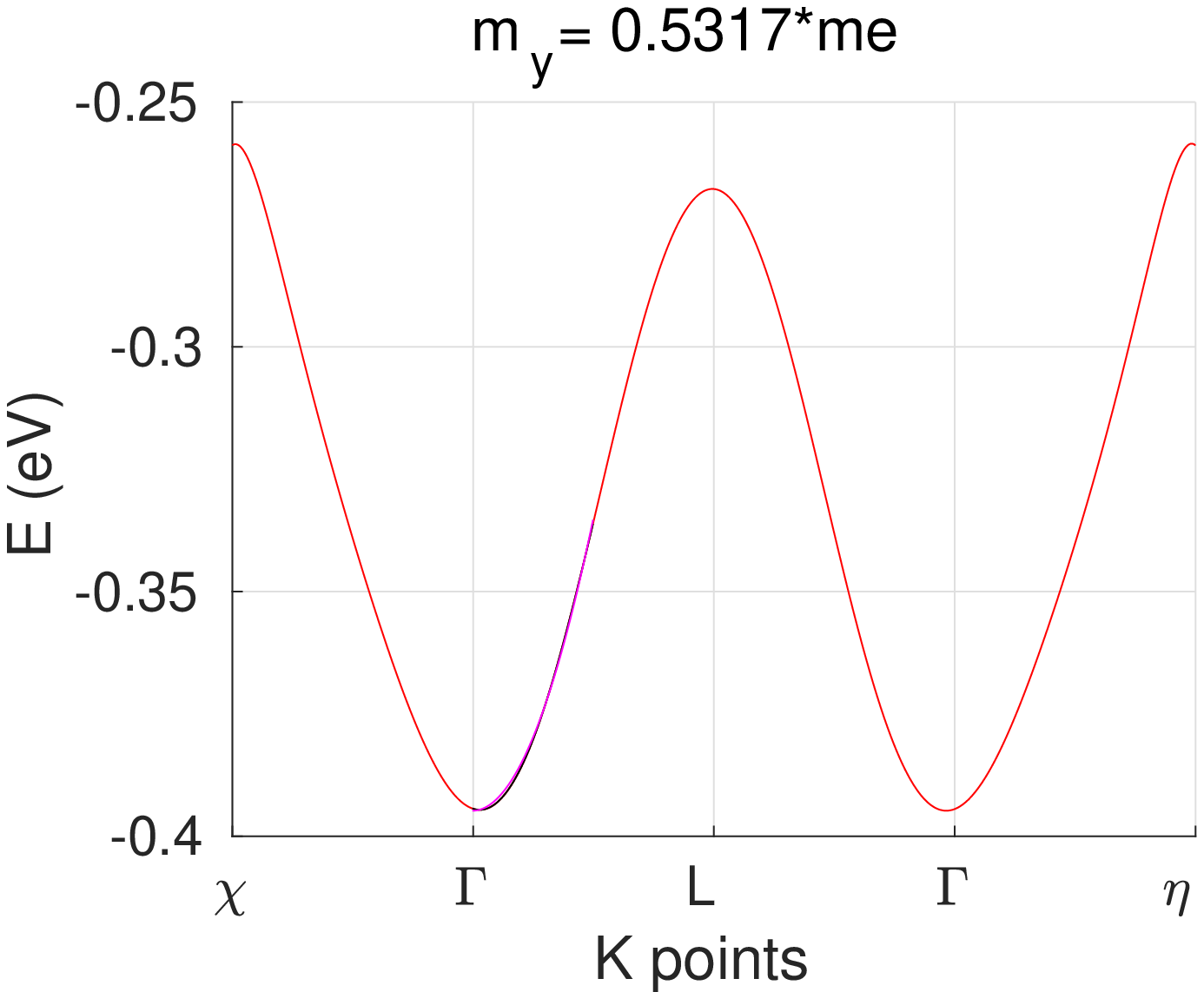}
    \caption{}
    \label{fig:eff_f2}
  \end{subfigure}
  \begin{subfigure}[b]{0.32\textwidth}
    \includegraphics[width=\textwidth]{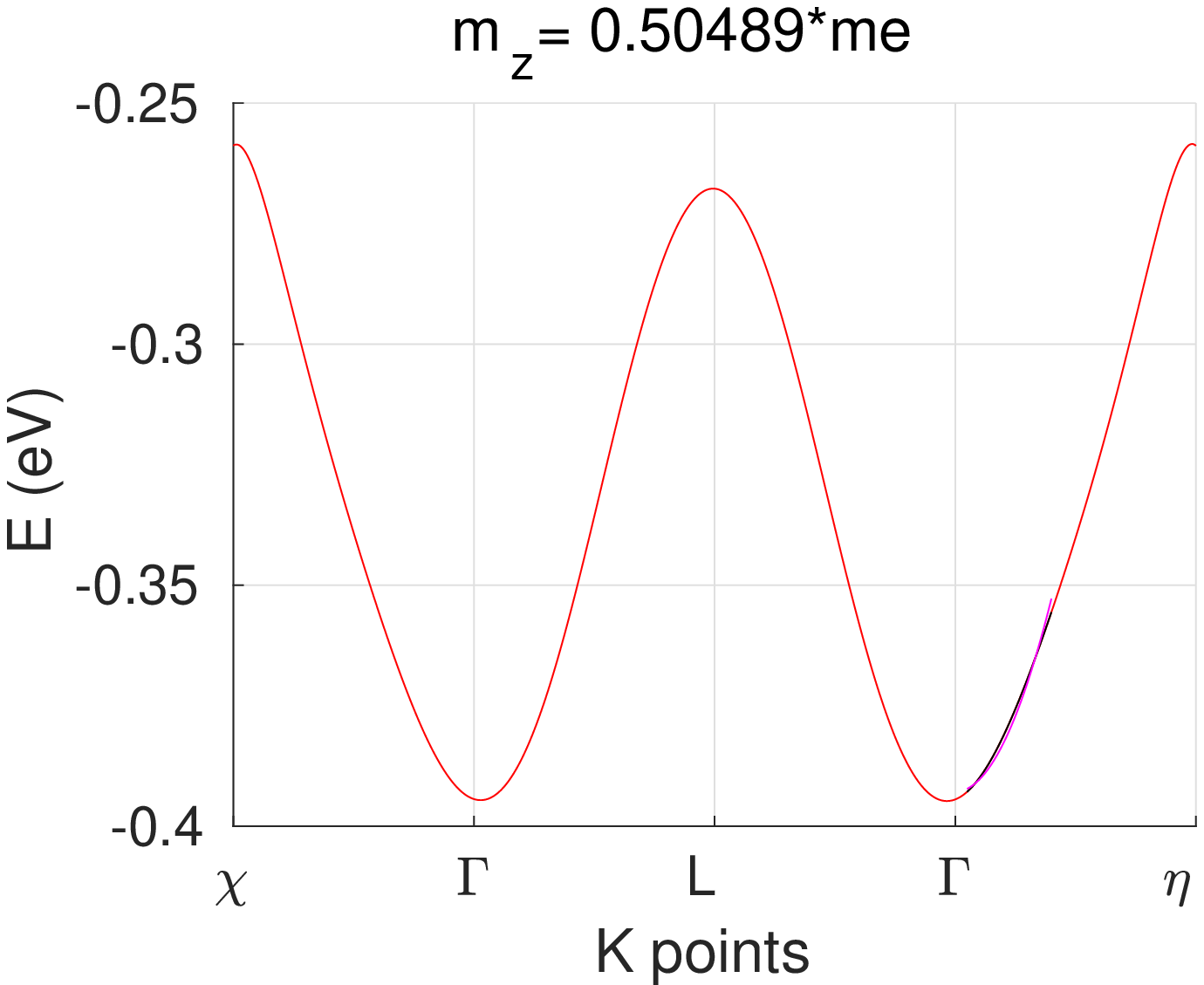}
    \caption{}
    \label{fig:eff_f3}
  \end{subfigure}
  \hfill
    \begin{subfigure}[b]{0.32\textwidth}
    \includegraphics[width=\textwidth]{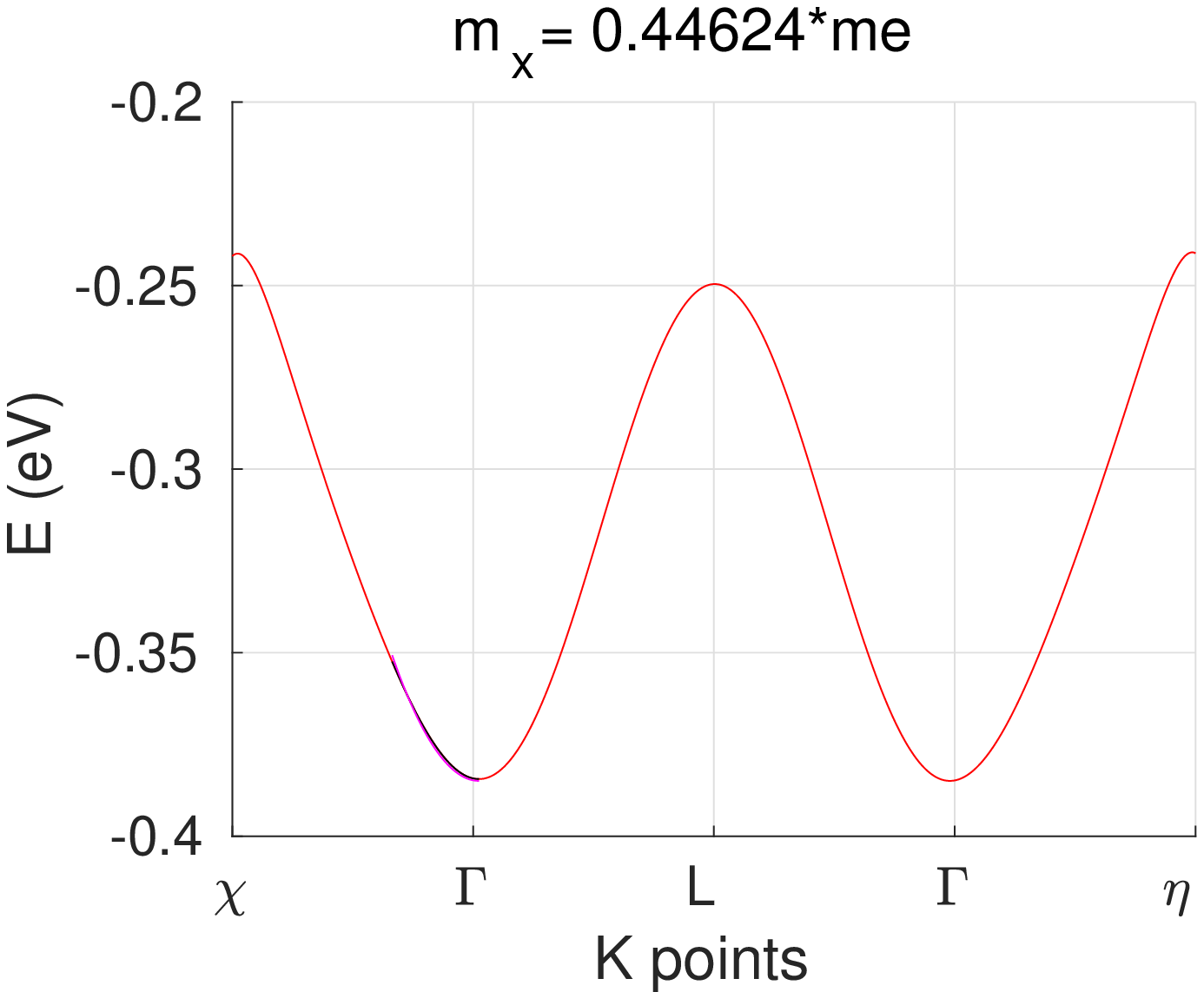}
    \caption{Spin up bands}
    \label{fig:effla_f1}
  \end{subfigure}
  \begin{subfigure}[b]{0.32\textwidth}
    \includegraphics[width=\textwidth]{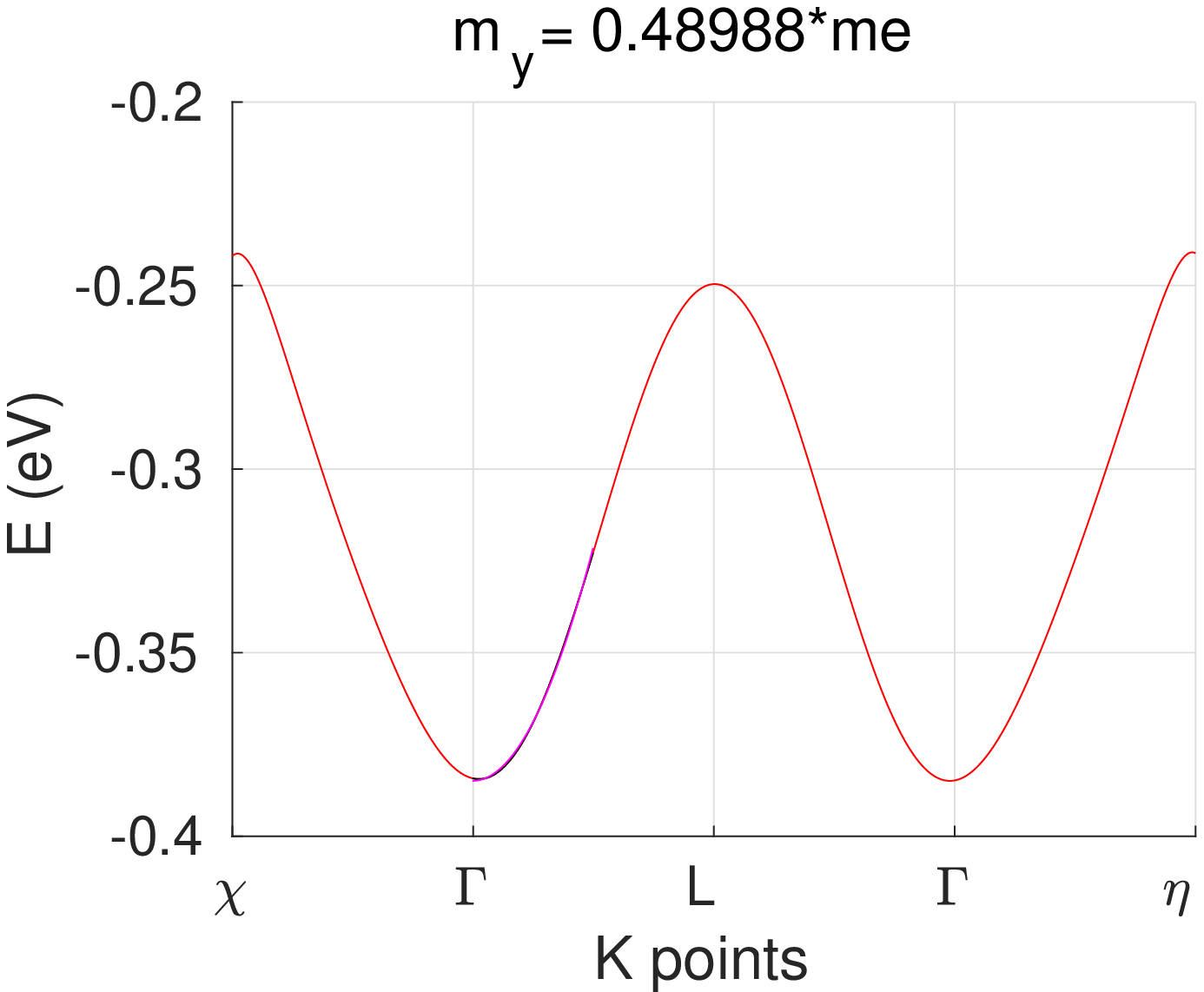}
    \caption{Spin down bands}
    \label{fig:effla_f2}
  \end{subfigure}
  \begin{subfigure}[b]{0.32\textwidth}
    \includegraphics[width=\textwidth]{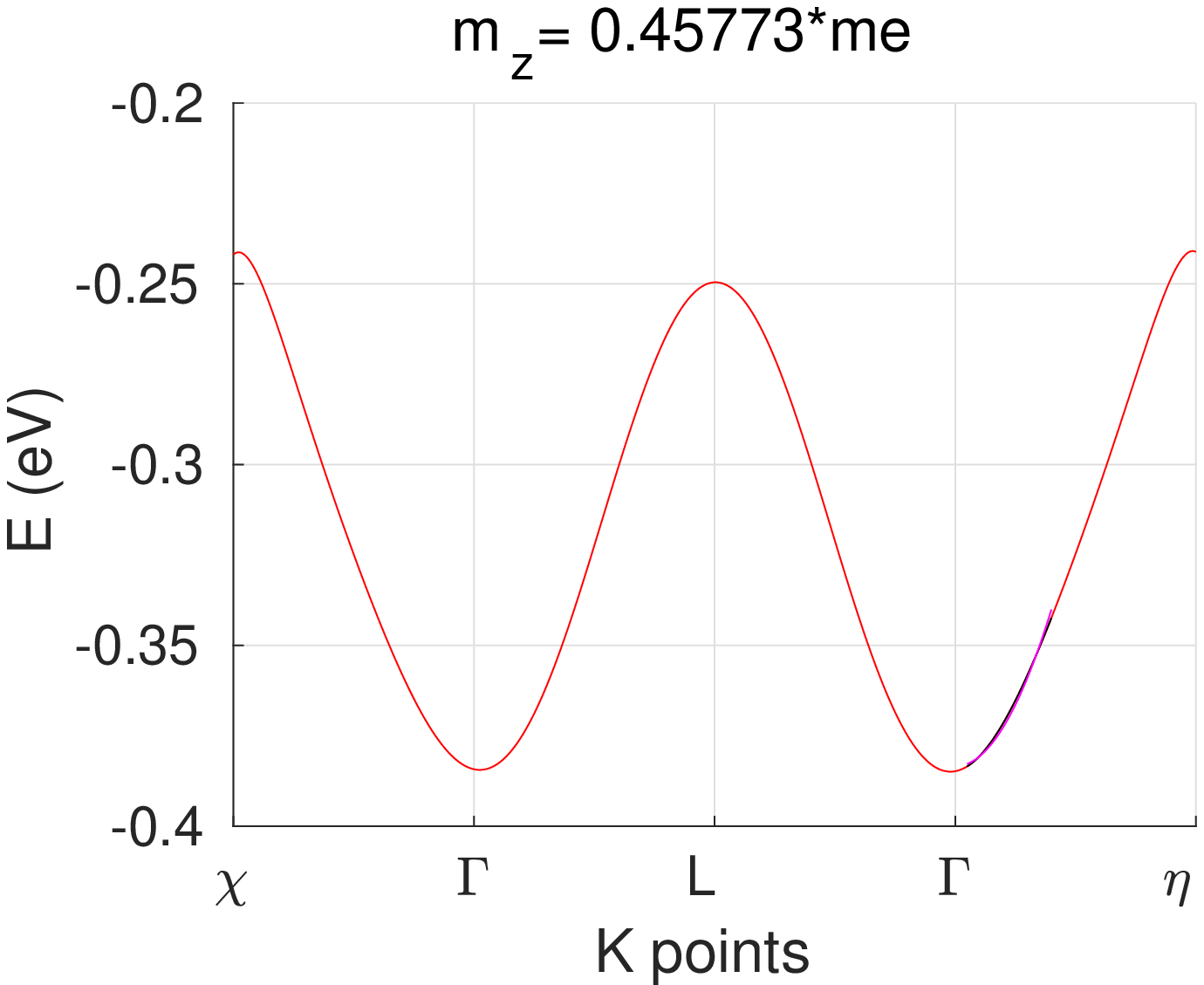}
    \caption{Combined bands}
    \label{fig:effla_f3}
  \end{subfigure}
  \captionof{figure}[123]{Fitting effective mass curves to the conduction band in the three coordinate directions. ~\ref{fig:eff_f1}: effective mass of conduction band in X to $\Gamma$ direction. ~\ref{fig:eff_f2}: effective mass of conduction band in $\Gamma$ to L direction. ~\ref{fig:eff_f3}: effective mass of conduction band in $\Gamma$ to $\eta$ direction.}
    \label{fig:YIG_eff}
\end{figure*}

The band gap of the YIG lattice calculated from the Ab intio calculations was 0.3262eV which is corroborated with values reported in the literature~\citep{0295xingtao}, where a band gap of 0.33eV was reported. As discussed in the previous section, each spin channel is treated independently and the effective mass of the spin down conduction band was calculated as shown in Figures~\ref{fig:eff_f1},~\ref{fig:eff_f2} and~\ref{fig:eff_f3}. The effective mass values in the three reciprocal vector directions $\chi-\Gamma$ (m$_x$),  $\Gamma$ to L (m$_y$) and $\Gamma$ to $\eta$ (m$_z$) are 0.4821*m$_e$, 0.5317*m$_e$ and 0.5048*m$_e$ respectively. The harmonic mean of the three effective mass values given in Equation~\ref{eq:effmass} gives the effective mass of the spin down conduction band to be m$_\downarrow$=0.5054*m$_e$, which is in close approximation to the value of 0.52*m$_e$ reported in the literature~\citep{0295xingtao}. Compared to the $\downarrow$ channel, the $\uparrow$ channel has a lower effective mass leading to its high mobility but as the energy gap of the $\uparrow$ conduction band from the Fermi region being high 1.1689 eV in comparison to 0.3262 eV of the YIG lattice, makes $\uparrow$ channel less conductive. Having the fundamental parameters of YIG, such as the lattice constant, band gap and effective mass match with the literature, a similar first principles approach is implemented for La:YIG.

\subsection{\label{sec:LaYIG_data}La:YIG Model Results}
The Y in the garnets are large cations having dodecahedral coordination. Substituting one Y atom with another large cation La yields LaY$_2$Fe$_5$O$_{12}$ (La:YIG). Figure~\ref{fig:UC_LaYIG} shows the super cell of La:YIG. Here 4 Y atoms are replaced with 4 La atoms and the relaxation of the unit cell is made following the description given in computational details section. As there are 12 Y positions in an 80-atom super cell, to replace 4 Y atoms with 4 La atoms there are 12C4 combinations. 12C4 yields 495, meaning 495 various unit cells must be optimized and the optimal cell that gives the least energy state must be chosen for further analysis. As this is computationally expensive, one means to perform this task is by mixing the pseudopotentials available in quantum espresso package. Mixing Y and La pseudopotentials can give freedom in choosing positions to replace Y with La. Following the mixing procedure embedded with quantum espresso package, a convergence in total energy and forces was achieved with the energy cutoff values the same as discussed in computational details section. The relaxed unit cell was used to obtain their electronic band structure, density of states, band gap, electron effective mass, the Fermi energy level, and the magnetization of the material. The lattice parameter increased by 0.87$\%$ compared to pure YIG due to La atoms being larger compared to Y and acquiring a value of 11.72 $\AA$. These properties were used to calculate its spin transport characteristics in the presence of a temperature gradient.

As shown in Figure~\ref{fig:YIG_bands}, both the $\uparrow$ and $\downarrow$ channels of La:YIG with band gaps of 1.1546eV (Figure~\ref{fig:laf1}) and 2.1143eV (Figure~\ref{fig:laf2}) respectively, saw a reduction in band gaps when compared with respective counter parts of YIG (Figures~\ref{fig:f1} and ~\ref{fig:f2}). But the overall band gap of the La:YIG material with 0.3544eV (Figure~\ref{fig:laf3}) has a slight increase in the band gap when compared to YIG material (Figure~\ref{fig:f3}). This makes each independent channel of La:YIG to be more conductive while the material is more insulating than YIG. Like YIG, La:YIG also has spin up channel to contribute to the valence band while spin down channel contributes to the conduction band. The effective mass values of La:YIG in the three reciprocal vector directions $\chi-\Gamma$ (m$_x$),  $\Gamma$ to L (m$_y$) and $\Gamma$ to $\eta$ (m$_z$) are 0.4462*m$_e$, 0.4898*m$_e$ and 0.4577*m$_e$ respectively. The harmonic mean of the three the values yields the effective mass of the $\downarrow$ conduction band to be 0.4639*m$_e$. The effective mass of La:YIG is less than that of YIG which has 0.5048*m$_e$, thus making the conduction electrons of La:YIG to have more mobility than YIG.

\subsection{\label{sec:validation}Validation of Model}
\begin{figure}[!ht]
\centering
\vspace{-1cm}
\includegraphics[trim={0 -1cm 0 -2cm},width=1\columnwidth]{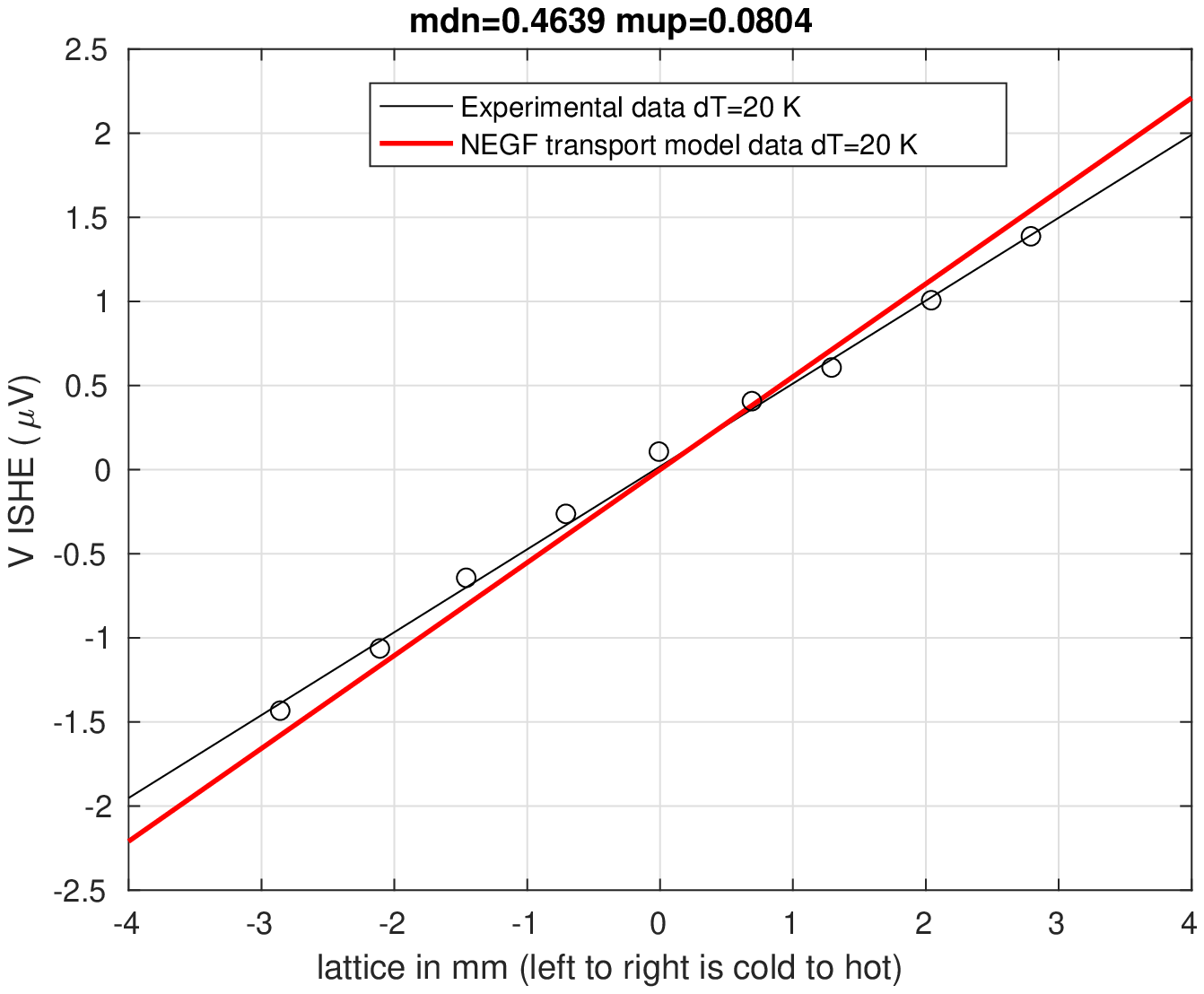}
\captionof{figure}[validation]{Validation of the model with the experimental data available for La:YIG. The $V_{ISHE}$ voltage in $\mu$V is calculated at the different points on the material lattice along the x-direction.}
\vspace{-0.5cm}
\label{fig:validation}
\end{figure}

After creating the Hamiltonian of the scattering region, a fine spectrum of energy bands in a range 0 to 5eV was created for each eigen energy level to calculate the total current in the region due to thermal bias. As the model treats the two spin channels independent, the relative movement of the spin up and spin down electrons is different and leads to a spin redistribution in the material due to the non-equilibrium induced by the contacts.

Using the fundamental parameters from~\ref{sec:LaYIG_data}, the approach described in the sections~\ref{sec:level2} and~\ref{sec:Spin_Seebeck} of this paper was implemented in MATLAB to calculate the spin voltage in the transversely attached Pt NM contact. To compare with the literature, Uchida et al~\citep{spinsemiconducting:1m}, a temperature bias of 20K is assumed through the analysis. The calculations were performed at 300K temperature (of the NM contact) and linear increase of temperature from left contact (cold) to the right contact (hot), Figure~\ref{fig:three_probe_system}. Table~\ref{tab:parameters} shows the experimental values and the parameters incorporated in the model. The model used the length of the scattering region along x-direction to be 36nm. A higher length can be incorporated depending on the computational power in reach. On an 8-core processor, a 36nm length scale took 4 hours to complete the calculation. The number of grid points, N=300, were chosen until the relative channel potential (U) reached a value 1E-6eV.
\begin{table}
\begin{center}
\caption{Parameters used in the experimental setup of Figure~\ref{fig:three_probe_system} as reported in literature~\citep{spinsemiconducting:1m} and the parameters used in verification.}
\begin{tabular}{ | c | c | c | c |}
\hline
{Material}  & \thead{Length \\ x-direction} & \thead{Length \\ y-direction} & \thead{Length \\ z-direction} \\
\hline
\thead{La:YIG \\ experimental} & {8 mm} &  \makecell{4 mm}  & {3.9 $\mu$m}  \\
\hline
\thead{Pt \\ experimental} & {0.1 mm} &  \makecell{4 mm}  & {15 nm}  \\
\hline
\thead{NEGF model \\ scattering region } & {36 nm} &  \makecell{18 nm}  & {17.5E-3 nm}  \\
\hline
\thead{NEGF model \\ NM contact} & {0.45 nm} &  \makecell{18 nm}  & {6.75E-5 nm}  \\
\hline
\end{tabular}
\label{tab:parameters}
\end{center}
\end{table}

The model incorporates the same ratio of L$_{x-dir}$:L$_{y-dir}$:L$_{z-dir}$ of Pt NM contact as used in experimental verification. The ratio being 1:40:15E-5. Here, the length of the Pt strip along y direction is related to the length of the La:YIG material along the y-direction which are the same values. Similarly, the length of FM material along the x-direction is related to the length of the NM contact along the y-direction. The ratio of the parameters in the model are same as those in the experimental setup. As a 1-D model was developed in this research, the quantities like spin current injected into the NM contact along the z-direction have the units A/m$^2$ and the population density of individual spin channels will have the units 1/m$^2$. Hence, the spin populations at each grid point along x-direction, Figure~\ref{fig:lattice}, were calculated and multiplied with the area of contact between NM (Pt) and FM (La:YIG), which is 0.45x18 nm$^2$. The difference in the populations of the spin channels causes the spin current to flow along the z-direction into the transversely NM contact. This spin current from La:YIG is converted into inverse spin Hall voltage V$_{ISHE}$ in the Pt contact. The other constants extracted from literature are shown in Table~\ref{tab:otherparameters}.

\begin{table}
\begin{center}
\caption{Constants incorporated in the model.}
\begin{tabular}{ | c | c |}
\hline
{Constant}  & \thead{Value} \\
\hline
\thead{$\gamma$ \\$^{reference(}$~\citep{Xiao:1y1}$^)$} & {1.4x10$^{11}$ 1/T.s} \\
\hline
\thead{$\rho$\\$^{reference(}$~\citep{UCHIDA2010524}$^)$\\ electrical resistivity of Pt at 300K} & {15.6E-8 $\Omega$.m} \\
\hline
\thead{g$_r$/A\\$^{reference(}$~\citep{grvaluelite}$^)$} & {0.1x10$^{16}$ 1/m$^2$} \\
\hline
\thead{ $\theta_H$\\$^{reference(}$~\citep{PhysRevhallangle}$^)$ \\Hall angle of Pt} & {0.0037}  \\
\hline
\thead{ V$_a$ \\ volume of NM} & {0.45x18x6.75E-5 nm$^3$}  \\
\hline
\end{tabular}
\label{tab:otherparameters}
\end{center}
\end{table} 

Incorporating the parameters from Tables~\ref{tab:parameters},~\ref{tab:otherparameters} in the aforementioned model, the V$_{ISHE}$ in the transversely attached Pt electrode at various locations along the scattering region is shown in Figure~\ref{fig:validation}. The red line represents the trend calculated from the NEGF model for a 20K temperature bias and the black line with the data points represented in black circles is the experimental trend for the same temperature obtained from literature~\citep{spinsemiconducting:1m}. The slope of the NEGF trend is close to the experimental data with an error of 6.5$\%$. The developed model can be applied to magnetic materials which have atoms with d-electrons. By assuming a $\pm$5$\%$ error in effective mass, the model predicts the curve with $\pm$15 $\%$ close to that of experimental value. Hence, this model can be applicable to materials whose experimental transverse spin properties are not available.

\section{\label{sec:Conclusion}Conclusions}
Through this research a 1-D model combining DFT, NEGF and spin transport theories was developed that treats both spin channels independently and calculates electronic and spin conductivities, and spin-Seebeck coefficient of the material. Validation was performed on La:YIG insulator material which proves the applicability of the model to various other semiconducting magnetic materials and opening a new avenue to theoretically evaluate the parameters that effect spin Seebeck coefficient. It is important to get converged results from DFT calculations as the accuracy in calculating the spin transport properties relies heavily on the material properties. In the case of La:YIG, a 3.2$\%$ error in calculating the inverse spin voltage was demonstrated with the potential to apply the model to much broader design spaces. Preliminary studied into other design spaces has demonstrated the subtle effects of substitutions in magnetic materials which control the spin polarization of the material can lead to significant improvements of the spin-thermoelectric performance.

\bibliography{bib_file}

\end{document}